\documentclass[fleqn,usenatbib]{mnras}

\usepackage[T1]{fontenc}

\usepackage{amstext,comment}
\usepackage{appendix}
\usepackage{graphicx}
\usepackage{amsmath,mathtools,revsymb,amssymb}
\usepackage{txfonts}
\usepackage{booktabs,multirow,calc,xspace,rotating}
\usepackage{tikz,xcolor,hyperref}

\definecolor{lime}{HTML}{A6CE39}
\DeclareRobustCommand{\orcidicon}{
	\begin{tikzpicture}
	\draw[lime, fill=lime] (0,0) 
	circle [radius=0.16] 
	node[white] {{\fontfamily{qag}\selectfont \tiny ID}};
	\draw[white, fill=white] (-0.0625,0.095) 
	circle [radius=0.007];
	\end{tikzpicture}
	\hspace{-2mm}
}

\foreach \x in {A, ..., Z}{\expandafter\xdef\csname orcid\x\endcsname{\noexpand\href{https://orcid.org/\csname orcidauthor\x\endcsname}
			{\noexpand\orcidicon}}
}

\title[Dust Eddington Ratios]{Dust Eddington Ratios for Star-Forming Galaxy Subregions}

\author[Blackstone \& Thompson]{Ian Blackstone$^{1,2}$\thanks{E-mail: blackstone.66@osu.edu} \orcidA \& Todd A.~Thompson$^{2,3,1}$\thanks{E-mail: thompson.1847@osu.edu} \orcidB \\
  $^{1}$Department of Physics, Ohio State University, 191 W. Woodruff Ave, Columbus, OH 43210 \\
  $^{2}$Center for Cosmology and Astro-Particle Physics, Ohio State University, 140 W.~18th Ave, Columbus, OH 43210 \\ 
  $^{3}$Department of Astronomy, Ohio State University, 140 W.~18th Ave, Columbus, OH 43210 \\
  }

\date{Accepted XXX. Received YYY; in original form ZZZ}

\pubyear{2023}

\begin{document}
\label{firstpage}
\pagerange{\pageref{firstpage}--\pageref{lastpage}}
\maketitle

\begin{abstract}
    Radiation pressure on dust is an important feedback process around star clusters and may eject gas from bright sub-regions in star-forming galaxies. The Eddington ratio has previously been constructed for galaxy-averaged observations, individual star clusters, and Galactic HII regions. Here we assess the role of radiation pressure in thousands of sub-regions across two local star-forming galaxies, NGC 6946 and NGC 5194. Using a model for the spectral energy distribution from stellar population synthesis and realistic dust grain scattering and absorption, we compute flux- and radiation pressure-mean opacities and population-averaged optical depth $\langle\tau_{\rm RP}\rangle$. Using Monte-Carlo calculations, we assess the momentum coupling through a dusty column to the stellar continuum. Optically-thin regions around young stellar populations are $30-50$ times super-Eddington. We calculate the Eddington ratio for the sub-regions including the local mass of young and old stars and HI and molecular gas. We compute the fraction of the total star formation that is currently super-Eddington, and provide an assessment of the role of radiation pressure in the dusty gas dynamics. Depending on the assumed height of the dusty gas and the age of the stellar population, we find that $\sim0-10$\% of the sightlines are super-Eddington. These regions may be accelerated to $\sim5-15$\,km/s by radiation pressure alone. Additionally, our results show that for beamed radiation the function $1-\exp(-\langle\tau_{\rm RP}\rangle)$ is an excellent approximation to the momentum transfer. Opacities and optical depths are tabulated for SEDs of different stellar ages and for continuous star formation. 
\end{abstract}


\section{Introduction}
\label{section:introduction}

Star clusters form inside Giant Molecular Clouds (GMCs),  collections of gas and dust that dominate the star formation in galaxies. Despite hosting as much as millions of solar masses worth of gas, only a relatively small percentage of the gas in these clouds may end up becoming stars \citep{Kennicutt1998,Krumholz2007}. This inefficiency in star formation in a region filled with the raw material for stars points to feedback mechanisms that inherently limit star formation \citep{Mckee2007,Murray2010b,Thompson2016}.

Many candidate feedback effects and mechanisms have been suggested in forming star clusters, including massive star supernovae, proto-stellar jets, stellar winds, ionized gas pressure, and radiation pressure on dust \citep{Mckee2007,Murray2005,Thompson2005, Krumholz2009,Draine2011,Lopez2011,Harwit1962,Odell1967,Tsang2015,Tsang2018,Menon2022b}. Recent simulations examine the combined and individual contributions of these feedback mechanisms (e.g., \citealt{Grudic2021}). Together with improvements in modern simulations, there is an effort to more fully assess theoretical models in light of extensive observations of both individual star clusters \citep{Lopez2011,Pellegrini2011} and HII regions \citep{Olivier2020}, GMCs \citep{Murray2010b}, and over galaxy scales \citep{Thompson2005,Andrews2011,Skinner2015,Wibking2018}. Similarly, with large galaxy-wide multi-wavelength datasets (e.g., PHANGS; \citealt{Leroy2019,Lee2022}), we can begin to ask whether specific feedback mechanisms can be demonstrably shown to dominate in specific environments.

In order to assess the importance of radiation pressure in galaxies, previous works have focused on comparing the observed flux to the dust Eddington limit, either on whole galaxy scales \citep{Murray2005,Thompson2005,Andrews2011,Coker2013,Wibking2018,Crocker2018a}, in external star-forming galaxy sub-regions \citep{Scoville2001,Krumholz2009,Murray2010b,Andrews2011}, or by examining local star clusters \citep{Lopez2011,Pellegrini2011} and individual HII regions \citep{Olivier2020,Pellegrini2011}. These works have examined both the ``single-scattering" Eddington limit, where the dust reprocessed FIR emission from absorption is not re-absorbed (sometimes called ``direct radiation pressure"), and the highly FIR optically-thick limit relevant for very dense ultra-luminous infrared galaxies and high-mass GMCs and proto-clusters \citep{Thompson2005,Krumholz2012,Krumholz2013,Davis2014,Zhang_Davis}. While the latter optically-thick limit is likely particularly relevant for forming super-star clusters as in \cite{Leroy2018,Levy2021}, the former single-scattering limit ``direct" radiation pressure has been argued to be most important for less extreme environments \citep{Murray2005,Krumholz2009,Murray2010b,Murray2011,Thompson2015,Skinner2015,Raskutti2016,Raskutti2017,Crocker2018a,Crocker2018b,Menon2022a}. Early studies of radiation pressure suggest that single-scattering may be the source of the acceleration of gas from star clusters \citep{Harwit1962}, leading to shell structures of dusty gas around stars \citep{Odell1967,Elmegreen1982}.

In the context of the Eddington limit for whole galaxies, several previous works have focused on comparing the estimates of the bolometric flux from star formation with the total average gas mass across the whole galaxy, which is dominated by the molecular phase.

For example, echoing earlier results by \cite{Andrews2011},  \cite{Wibking2018} find that galaxies as a whole are significantly sub-Eddington when comparing the total bolometric flux with the Eddington flux $F_{\rm Edd}\sim G\Sigma_{\rm tot}\Sigma_{\rm gas} c$, where $\Sigma_{\rm tot}$ is the total surface density and $\Sigma_{\rm gas}$ is the {\it total} gas surface density, including the molecular phase. On the other hand, \cite{Lopez2011} find that direct radiation pressure can be significant on the scale of individual star clusters. Meanwhile, \cite{Thompson_Krumholz} argue that because of the form of the Eddington flux in the single-scattering limit -- $F_{\rm Edd}\propto \Sigma_{\rm tot}\Sigma_{\rm gas} $ -- low column-density sightlines will be preferably ejected from a super-sonically turbulent medium, and that super-Eddington sightlines exist despite the fact that the region is globally sub-Eddington.

These considerations motivate a new assessment of the role of radiation pressure on dusty gas. In this paper we provide a qualitatively different look at the importance of radiation pressure on dust in galaxies than previous discussions in the literature. In particular, motivated by the scaling of the Eddington flux with column density, we focus on the relatively low column density material along the line of sight to H$_\alpha$-emitting regions in the local star-forming galaxies NGC 5194 and NGC 6946. The data are taken from the work of \cite{Kessler2020}, who provide H$\alpha$ flux, extinction, and 3.6\,$\mu$m flux for several thousand $\sim40$\,pc-scale sub-regions across both galaxies. Instead of comparing the Eddington flux using the entire observed column, which would include the dense molecular gas, we instead ask if the observed regions are super/sub-Eddington using the extinction as measured in each sub-region, along the line of sight, as a proxy for the projected gas surface density. This is a change from most previous approaches, and is closer to the work of \cite{Murray2005}, but resolved across the face of galaxies. In addition to this change in tack from some earlier works, we also employ a more detailed treatment of the momentum coupling between the radiation field and the dusty gas. Instead of assuming a single constant value for the radiation pressure opacity, we use detailed dust properties, a realistic dust composition mixture, and include both anisotropic and multiple scattering \citep{Draine1984,Laor1993}. For our model of the photon field, we use BPASS simulation results for binary star populations at different ages to convert the observed H$\alpha$ luminosity to a bolometric luminosity and stellar mass in new stars \citep{Stanway2018,Eldridge2017}. Additionally, we construct Monte Carlo simulations to measure the momentum coupling between the stellar population SEDs and an overlying dusty column, including the wavelength-dependent scattering albedo, multiple scattering, and anisotropic scattering. This more detailed approach allows us to assess some of the commonly used analytic estimates for the momentum coupling between stellar populations and dusty gas in the semi-transparent regime.

In Section \ref{section:DustOpacitiesEddingtonRatios}, we discuss the BPASS models and the flux-mean and radiation pressure-mean dust opacities for our fiducial dust mixture. We compare  the optically-thin Eddington luminosity limit per unit mass  $L_{\rm Edd}/M$ to $L/M$ for the BPASS models with different assumed initial mass functions. Section \ref{section:MonteCarloResults} presents the Monte Carlo simulations performed to track momentum deposition in the dusty column, given the BPASS SEDs and the assumed dust grain properties. Wavelength-dependent anisotropic scattering is used to provide more physically accurate results for the simulation. We also provide a comparison with commonly-used analytic approximations. In Section \ref{section:ApplicationToGalaxies}, we discuss the data employed. We then apply our derived dust opacities to determine under what assumptions the observed star-forming sub-regions are sub/super-Eddington. For super-Eddington regions, we discuss a simple model of the shell dynamics, predicting velocity as a function of time.

Absent other feedback mechanisms, dusty gas surrounding super-Eddington sub-regions may be driven to a velocity large enough to expel it from the disk so that it ``sees" the whole galaxy, instead of just the initial driving region \citep{Murray2011,Hopkins2012}. The dynamics is dominated by the luminosity of the region, its escape velocity, and the density of the old stellar population, which contributes to the enclosed mass as the shell is driven outwards. 

Section \ref{section:discussion} provides a discussion and conclusion. Overall, we find that the question of whether or not a given region is sub/super-Eddington depends sensitively on the vertical height of the dusty gas along the line of sight, because this scale controls the total mass enclosed by the region, which we find is dominated by the old stellar population for the galaxies investigated here.

\section{Dust Opacities and Eddington Ratios for Stellar Populations}
\label{section:DustOpacitiesEddingtonRatios}

For a given IMF, BPASS provides broadband spectra for the stellar population as a function of age. For our fiducial case, we use a standard BPASS IMF with a high-mass Salpeter slope ($dN/dM\propto M^{-2.3}$) up to 300 $M_\odot$ and a metallicity of $z = 0.020$ evolved from 1\,Myr to 10\,Gyr \citep{Eldridge2017,Stanway2018}. Other assumptions about the IMF are discussed further below.

We adopt the notation for the IMF models used by \cite{Stanway2018}, with Equation \ref{equation:IMF} defining the mass distribution of stars. The mass range for these models goes from $0.1\,\textrm{M}_\odot$ to $M_{\rm max}=300\,\textrm{M}_\odot$ or $100\,\textrm{M}_\odot$. The definitions of $\alpha_1$ and $\alpha_2$, as well as the maximum mass for each model used in this paper can be found in Table \ref{table:IMFModels}, with the distribution given by
\begin{equation}
    N_{(\rm M<M_{\rm max})} \propto \int_{0.1}^{\rm M_1} \bigg(\frac{\rm M}{\rm M_{\odot}}\bigg)^{\alpha_1} d\textrm{M} + \textrm{M}_1^{\alpha_1}\int_{\rm M_1}^{\rm M_{\rm max}} \bigg(\frac{\rm M}{\rm M_{\odot}}\bigg)^{\alpha_2} d\textrm{M}.
    \label{equation:IMF}
\end{equation}
The fiducial model high-mass IMF is a Salpeter IMF, but a flatter IMF model, as motivated by \cite{Schneider2018}, is also tested. The choice of IMF model sets the SED and L/M for the ZAMS population, as well as its time evolution.
\begin{table}
\centering
\begin{tabular}{ccccc}
Model & $\alpha_1$ & $\alpha_2$ & $\rm M_1 \, (M_\odot)$ & $\rm M_{\rm max} \, (M_\odot)$ \\ \hline
135\_300 & -1.30 & -2.35 & 0.5 & 300 \\
135\_100 & -1.30 & -2.35 & 0.5 & 100 \\
100\_300 & -1.30 & -2.00 & 0.5 & 300 \\
Chab300 & exp cutoff & -2.3 & 1 & 300 
\end{tabular}
\caption{Subset of IMF models from \protect\cite{Stanway2018} used in this paper. Equation \protect\ref{equation:IMF} provides the IMF for each model given the parameters. Our fiducial model in this paper is the model 135\_300, which uses a Salpeter high-mass IMF slope. Other IMFs are discussed.} 
\label{table:IMFModels}
\end{table}

We use precalcuated values of the grain properties \citep{Draine1984,Laor1993}\footnote{\url{https://www.astro.princeton.edu/~draine/dust/dust.diel.html}}. We interpolate the dust data to the wavelengths in the BPASS spectral models and using these we calculate grain size-averaged and spectrum-averaged cross sections for the grain types. Specifically, we calculate the radiation pressure mean opacity and the flux-mean opacity as
\begin{equation}
    \langle \kappa_{\rm RP} \rangle = f_{\rm dg} \frac{\int \pi a^2 \frac{dn}{da} \int (Q_{\rm abs, (a,\,\lambda)} + (1-g_{(a,\,\lambda)})Q_{\rm scatt,(a,\,\lambda)})L_{\lambda} d\lambda da}{L_{\rm bol}\int \frac{dn}{da} \frac{4\pi}{3} \rho_{\rm grain} a^3 da},
    \label{equation:kappa_av_rp}
\end{equation}
and
\begin{equation}
    \langle \kappa_{\rm F} \rangle = f_{\rm dg} \frac{\int \pi a^2 \frac{dn}{da} \int (Q_{\rm abs, (a,\,\lambda)} + Q_{\rm scatt, (a,\,\lambda)})\,L_{\lambda} d\lambda da}{L_{\rm bol}\int \frac{dn}{da} \frac{4\pi}{3} \rho_{\rm grain} a^3 da}.
    \label{equation:kappa_av_f}
\end{equation}
For the purposes of our Monte Carlo calculations described in the next section it is also useful to note the wavelength dependent versions of these same relations
\begin{equation}
    \kappa_{\rm RP, (\lambda)} = f_{\rm dg} \frac{\int \pi a^2 \frac{dn}{da} (Q_{\rm abs, (a,\,\lambda)} + (1-g_{(a,\,\lambda)})Q_{\rm scatt,(a,\,\lambda)}) da}{\int \frac{dn}{da} \frac{4\pi}{3} \rho_{\rm grain} a^3 da},
    \label{equation:kappa_rp}
\end{equation}
and
\begin{equation}
    \kappa_{\rm F, (\lambda)} = f_{\rm dg} \frac{\int \pi a^2 \frac{dn}{da} (Q_{\rm abs, (a,\,\lambda)} + Q_{\rm scatt,(a,\,\lambda)}) da}{\int \frac{dn}{da} \frac{4\pi}{3} \rho_{\rm grain} a^3 da}.
    \label{equation:kappa_f}
\end{equation}
In the above expressions, $Q_{\rm abs}$ and $Q_{\rm scatt}$ are the wavelength-dependent absorption and scattering efficiencies, respectively, $L_{\rm bol}$ is the integrated bolometric luminosity of the stellar population across the BPASS SED, $\rho_{\rm grain}$ is the dust grain mass density, $a$ is the grain size, $g \equiv \langle \cos(\theta) \rangle$ is the average wavelength dependent scattering angle supplied by the grain data files \citep{Draine1984,Draine2011}. For the grain size distribution, $dn/da$, we use the MRN distribution \citep{Mathis1977}, $dn/da\propto a^{-3.5}$. We assume the grains are spheres of radius $a$ with grain mass density $\rho_{\rm grain}$. For our fiducial model, we use a maximum grain size $a_{\rm max} = 1 \, \mu$m and a minimum grain size $a_{\rm min} = 0.001\, \mu$m and we calculate these opacities for silicon-carbide, astronomical silicate, and amorphous carbon. We use $\rho_{\rm grain,\,SiC} = 3.22 \,\textrm{g/cm}^3$ for Silicon-Carbide, $\rho_{\rm grain,\,Sil} = 3.3 \,\textrm{g/cm}^3$ for Silicate, and $\rho_{\rm grain,\,Gra} = 2.26 \,\textrm{g/cm}^3$ for Carbon \citep{Laor1993}. The dust is composed of a mixture of $50\%$ astronomical silicates, $45\%$ amorphous carbon, and $5\%$ silicon-carbide. The ratios that make up our fiducial dust model are taken from \cite{Laor1993}. Throughout this paper we normalize to a total dust-to-gas mass ratio $f_{\rm dg}=1/100$. Variations in this ratio are discussed in Section \ref{section:discussion}.

\begin{figure*}
\includegraphics[width=0.95\textwidth]{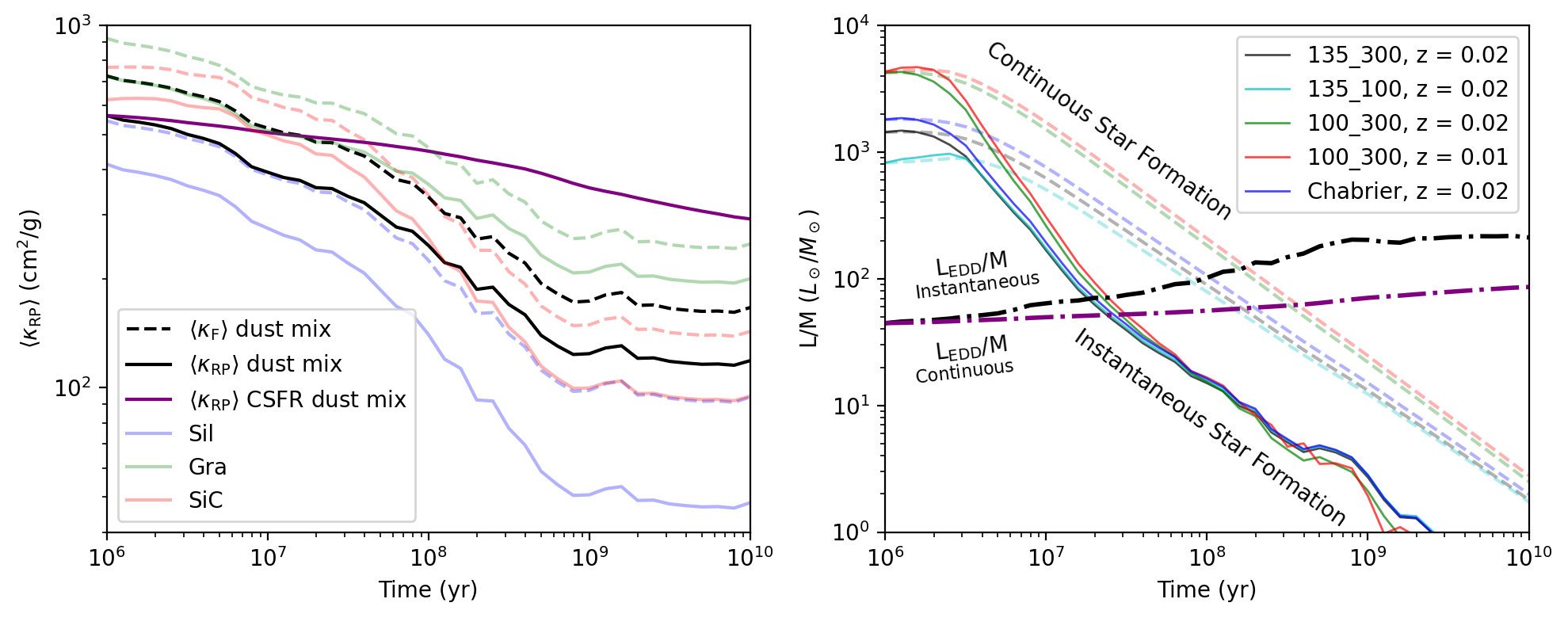}
\caption{Left: Opacity for different grain types for a 135\_300 solar metallicity population of instantaneously-formed stars over time. As the stars age the population reddens, which causes the opacity to decrease as a function of time. Right: Luminosity to mass $L/M$ ratio for instantaneously and continuously formed stars using the BPASS binary synthesis calculation for various initial IMFs and metallicities. The dash-dotted lines are the optically thin $\rm L_{\rm Edd}/M$ ratio for our fiducial dust model using both an instantaneous formed and a continuously formed stellar population. An optically-thin dusty atmosphere is significantly super-Eddington for the first $\sim30$ Myr of the lifetime of a stellar population. Continuous star formation in optically thin regions is super-Eddington beyond 100 Myr \citep{Eldridge2017,Stanway2018}.}
\label{figure:KappaAndLOverM}
\end{figure*}

\begin{figure}
    \includegraphics[width = 0.43\textwidth]{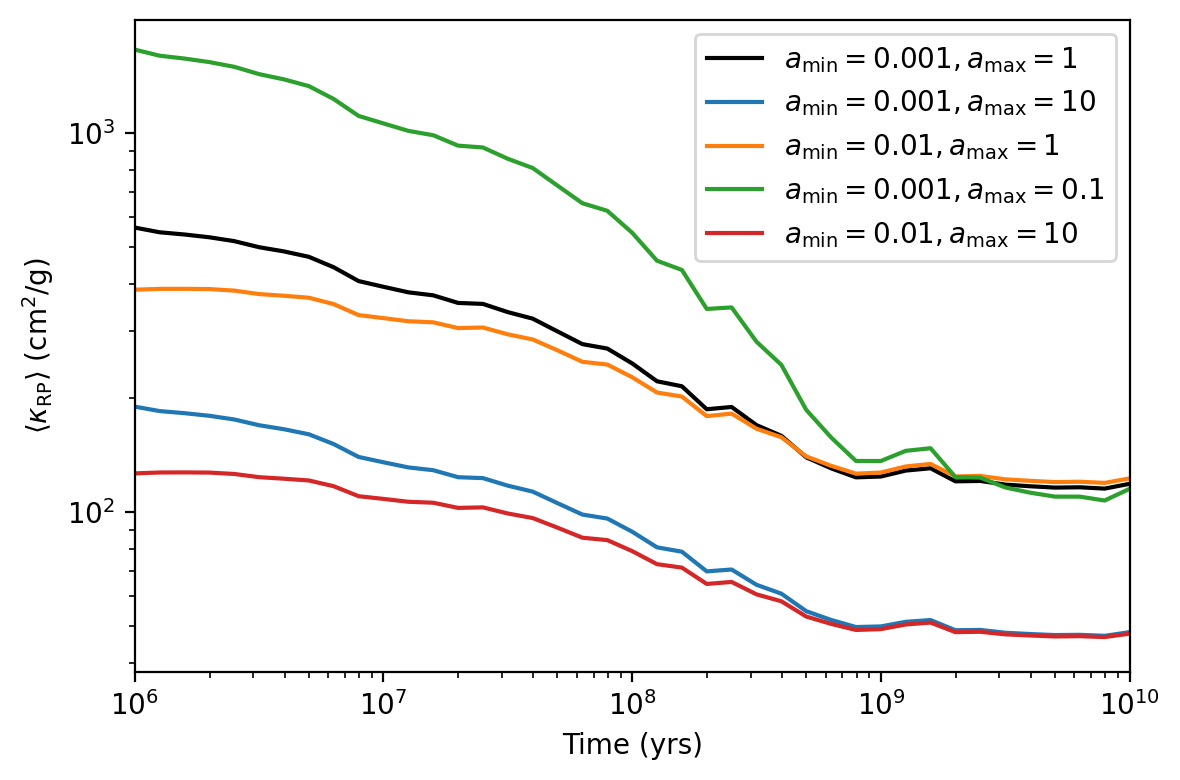}
    \caption{Values for $\langle\kappa_{\rm RP}\rangle$ for different assumptions of the maximum and minimum grain size in the grain distribution. The fiducial model is the black line with $a_{\rm min} =  0.001\,\mu\rm{m}$ to $a_{\rm max} =  1\,\mu\rm{m}$.}
    \label{figure:grain_size_comparison}
\end{figure}

Using these inputs, the left panel of Figure \ref{figure:KappaAndLOverM} shows calculations of the radiation pressure and flux-mean dust opacities (eqs.~\ref{equation:kappa_av_rp} and \ref{equation:kappa_av_f}) as a function of time for the fiducial BPASS 135\_300 stellar population (see Equation \ref{equation:IMF}; Table \ref{table:IMFModels}). Our dust mixture is shown as the black lines. The purple line shows $\langle\kappa_{\rm RP}\rangle$ over time for the dust mixture in a population with continual star formation. The contributions from the different grain species are shown as the other colored lines. We see that $\langle\kappa_{\rm RP}\rangle$ and $\langle\kappa_{\rm F}\rangle$ are $\simeq500$ and $700$\,cm$^2$ g$^{-1}$, respectively, for young $\sim$\,Myr-old stellar populations. As the stellar population ages and reddens, the spectrum-averaged opacities for the instantaneous star formation models decrease by about a factor of 4 on Gyr timescales to $\langle\kappa_{\rm RP}\rangle\simeq100$ and $\langle\kappa_{\rm F}\rangle\simeq150$\,cm$^2$ g$^{-1}$. These results are reported for an assumed dust-to-gas total mass ratio $f_{\rm dg}$ of $1/100$. Figure \ref{figure:grain_size_comparison} shows the calculated $\langle\kappa_{\rm RP}\rangle$ over time for different choices of $a_{\rm max}$ and $a_{\rm min}$ using our fiducial spectra. It shows that the choice of minimum and maximum grain size have a large impact for young stellar clusters, but the behavior for populations older than $1\,$Gyr is largely determined by the choice of the maximum grain size. At all ages the opacity scales with changes to the maximum grain size as $\approx 1/\sqrt{a_{\rm max}}$. Scaling from changes in the minimum grain size, however, are less impactful and change over time as the SED changes. The effects of grain size distribution and other details of the grain model are further discussed in Appendix \hyperref[appendix:DustGrains]{A}.

We wish to compare the  luminosity-to-mass ratio for BPASS stellar populations with the Eddington limit for an optically-thin dusty column. In an optically-thin medium, the Eddington limit is
\begin{equation}
    \frac{L_{\rm Edd}}{M_{\rm tot}}=\frac{4\pi G c}{\langle \kappa_{\rm RP}\rangle} \approx 50 \, \frac{L_\odot}{M_\odot}\bigg(\frac{500 \,\textrm{cm}^2\textrm{g}^{-1}}{\langle \kappa_{\rm RP} \rangle}\bigg),
    \label{equation:L_Edd_op_thin}
\end{equation}
where in the second approximate equality we normalize to a value of $\langle\kappa_{\rm RP}\rangle$ appropriate to a young stellar population shown in Figure \ref{figure:KappaAndLOverM} (left panel). The right panel of Figure \ref{figure:KappaAndLOverM} shows $L/M$ for our fiducial IMF model and instantaneous star formation, for the IMF models 135\_100 and 100\_300 at two different metallicities, and the Chabrier IMF \citep{Chabrier2003} (see eq.~\ref{equation:IMF} and Table \ref{table:IMFModels}). For comparison, we also show $L/M$ under the assumption of continuous star formation (dashed black) for these same IMF models. The Eddington luminosity-to-mass ratio $L_{\rm Edd}/M$ for an optically-thin dust mixture is shown as the black and purple dash-dotted lines, for instantaneous and continuous star formation respectively, as calculated using Equation \ref{equation:L_Edd_op_thin} using our fiducial IMF and instantaneous star formation. Note that the values for $L_{\rm Edd}/M$ would be slightly different for any of the IMFs (or metallicities) in the limit of instantaneous star formation because each has a different SED.

The right panel of Figure \ref{figure:KappaAndLOverM} shows that in the optically-thin limit every IMF model is significantly super-Eddington for $\sim30$\,Myr. As the population reddens and ages, the Eddington luminosity per mass increases in the optically-thin limit because the SED-averaged radiation pressure opacity decreases (left panel). The system eventually becomes sub-Eddington because the luminosity decreases as the massive stars die. For continuous star formation we see that the population remains super-Eddington out to $\sim200-400$\,Myr, depending on the IMF. Note that in these plots we assume that the mass of the stellar population is not effected by supernovae and mass loss through stellar evolution. For an instantaneously formed stellar population with typical assumptions about neutron star masses, black hole formation rates \citep{Pejcha2015}, and mass loss from stars, we expect that the total mass of the system decreases by a factor of $\approx 2$ as the population ages. Thus, in Figure \ref{figure:KappaAndLOverM} we underestimate $L/M$ somewhat relative to $L_{\rm Edd}/M$, making instantaneously formed stellar populations more super-Eddington in the optically-thin limit.

\section{Momentum transfer efficiency from Monte Carlo Simulations}
\label{section:MonteCarloResults}

The results in the right panel of Figure \ref{figure:KappaAndLOverM} apply only for optically-thin dusty columns. As the column density of the medium exceeds $\sim\langle\kappa_{\rm RP}\rangle^{-1}$ the medium approaches the so-called ``single-scattering" limit where it is optically-thick to the incoming UV/optical continuum from the stellar population, but optically-thin to the far-infrared emission produced by absorption. For typical values, this critical gas column density is 
\begin{equation}
    \Sigma_{\rm gas}({\langle\tau_{\rm RP}\rangle=1})\simeq\langle\kappa_{\rm RP}\rangle^{-1}\simeq9.6\,{\rm M_\odot\,\,pc^{-2}}\left(\frac{500\,{\rm cm^2\,\,g^{-1}}}{\langle\kappa_{\rm RP}\rangle}\right),
    \label{equation:CriticalSigma}
\end{equation}
which correspond to $A_V\simeq 0.542$ and $A_{\rm H \alpha}\simeq 0.479$ in our fiducial dust model. For surface densities above these values, the Eddington limit changes.

We wish to extend our analysis to optically-thick columns in the single-scattering limit. A typical approximation to the momentum transferred to such an atmosphere is done by assuming that all photons that enter the atmosphere are absorbed, leading to a fractional momentum transfer of $1$. However, one expects a more detailed calculation to yield results different from unity, even for idealized planar geometry because of multiple scattering and anisotropic scattering.

To compute the momentum transfer more completely, we use a Monte Carlo simulation to track anisotropic Henyey-Greenstein \citep{Henyey1941} scattering from a source of photons in plane-parallel geometry. For the simulation, we define a reference photon wavelength and an optical depth $\tau_{\rm ref}$ for our reference photon. For this paper, we choose the reference wavelength to be H$\alpha$, $\lambda_{\rm ref} = 656$\,nm. The optical depth of a cloud is dependent on the surface mass density, which can be calculated using Equation \ref{equation:Sigma_g} derived from \cite{DrainTextbook2011}.
\begin{equation}
    \Sigma_{\rm gas} = \frac{\rm A_{\rm H\alpha}}{1.086 \kappa_{\rm F, H\alpha}}.
    \label{equation:Sigma_g}
\end{equation}
Here $A_{\rm H\alpha}$ is the observed H$\alpha$ extinction and $\kappa_{\rm F,\, H\alpha}$ is the H$\alpha$ flux opacity for our dust model calculated from Equation \ref{equation:kappa_f}. We can then find the reference optical depth
\begin{equation}
    \tau_{\rm RP,\, ref} = \Sigma_{\rm gas}\kappa_{\rm RP, ref},
    \label{equation:tau_RP_ref}
\end{equation}
where $\kappa_{\rm RP,\, ref}$ is calculated with Equation \ref{equation:kappa_rp}. We then sample photons from the BPASS SED for a stellar population of a given age and calculate the effective optical depth for each photon wavelength using
\begin{equation}
    \tau_{\rm RP,\, (\lambda)} = \tau_{\rm RP, \, ref}\frac{\kappa_{\rm RP,(\lambda)}}{\kappa_{\rm RP,\, ref}}.
    \label{equation:tau_rp_lambda}
\end{equation}
Each photon is given an initial upward direction in plane-parallel geometry. Our simulations consider  both uniformly-sampled upward angles from $ 0 \leq \mu = \cos(\theta) \leq 1$ (called ``isotropic") and ``beamed" photon sources with $\mu = 1$. These photons are then released into an atmosphere consisting of uniform gas and dust. To speed up the simulation and to reduce noise in the results, a single MRN averaged value is used for both the wavelength-dependent albedo,
\begin{equation}
    \langle \rm \omega \rangle_{(\lambda)} = \frac{\int a^{2}\frac{dn}{da} Q_{\rm scatt, (a,\lambda)} da}{\int a^{2}\frac{dn}{da} Q_{\rm scatt, (a,\lambda)} da + \int a^{2}\frac{dn}{da} Q_{\rm abs, (a,\lambda)} da},
    \label{equation:albedo_mean}
\end{equation}
and the wavelength-dependent scattering angle,
\begin{equation}
    \langle g \rangle_{(\lambda)} = \frac{\int \frac{dn}{da} g_{(a,\lambda)} da}{\int \frac{dn}{da} da}.
    \label{equation:g_mean}
\end{equation}
As photons absorb or scatter, the momentum transfer to the gas is tracked and over multiple runs at different optical depths we build up a picture of how momentum is deposited in an atmosphere.

Because each photon has its own value of $\tau_{\rm RP,\,(\lambda)}$, our results for the fractional momentum transfer and Eddington flux or luminosity for a stellar population of a given IMF and age will be plotted as a function of the spectrum-average optical depth,
\begin{equation}
    \langle\tau_{\rm RP}\rangle = \Sigma_{\rm gas}\langle\kappa_{\rm RP}\rangle,
    \label{equation:tau_av_RP}
\end{equation}
where $\Sigma_{\rm gas}$ is given by Equation \ref{equation:Sigma_g} and $\langle\kappa_{\rm RP}\rangle$ is given by Equation \ref{equation:kappa_av_rp}, with values for our fiducial model given in Table \ref{table:MeanOpacities}. Because the value of $\langle\tau_{\rm RP}\rangle$ for a given gas surface density $\Sigma_g$ of the atmosphere is dependent on the SED of the stellar cluster we note that for plots using $\langle\tau_{\rm RP}\rangle$ such as Figure \ref{figure:monte_carlo_result} (discussed below), each line is plotted against the $\langle\tau_{\rm RP}\rangle$ calculated for that stellar population's age and IMF.

\begin{table}
\centering
\begin{tabular}{ccccc}
 & \multicolumn{2}{c}{Instantaneous} & \multicolumn{2}{c}{Continuous}\\
$\log_{10}\,$(years) & $\langle\kappa_{\rm RP}\rangle $ & $\langle\kappa_{\rm F}\rangle $ & $\langle\kappa_{\rm RP}\rangle $ & $\langle\kappa_{\rm F}\rangle $\\
 & ($\textrm{cm}^2/\textrm{g}$) & ($\textrm{cm}^2/\textrm{g}$) & ($\textrm{cm}^2/\textrm{g}$) & ($\textrm{cm}^2/\textrm{g}$) \\\hline\hline 
6.0 & 563 & 724 & 563 & 724 \\
6.2 & 540 & 695 & 556 & 715 \\
6.4 & 519 & 670 & 545 & 703 \\
6.6 & 487 & 635 & 533 & 688 \\
6.8 & 442 & 580 & 520 & 673 \\
7.0 & 393 & 520 & 506 & 656 \\
7.2 & 373 & 497 & 495 & 643 \\
7.4 & 354 & 474 & 485 & 631 \\
7.6 & 324 & 436 & 475 & 619 \\
7.8 & 277 & 375 & 463 & 603 \\
8.0 & 246 & 336 & 449 & 587 \\
8.2 & 214 & 294 & 433 & 567 \\
8.4 & 189 & 261 & 417 & 547 \\
8.6 & 159 & 220 & 401 & 527 \\
8.8 & 130 & 181 & 378 & 498 \\
9.0 & 124 & 173 & 356 & 469 \\
9.2 & 130 & 183 & 341 & 450 \\
9.4 & 120 & 169 & 326 & 431 \\
9.6 & 117 & 163 & 313 & 414 \\
9.8 & 116 & 162 & 301 & 399 \\
10.0 & 118 & 166 & 292 & 387 \\ \hline\hline
\end{tabular}
\caption{The values for $\langle\kappa\rangle$ for both radiation pressure (eq. ~\ref{equation:kappa_av_rp}) and flux means (eq.~\ref{equation:kappa_av_f}) over time for the fiducial 135\_300 model for both instantaneous star formation and continuous star formation. Values for $\langle\kappa_{\rm RP}\rangle$ and $\langle\kappa_{\rm F}\rangle$ are per gram of gas, using our fiducial dust-to-gas ratio of $f_{\rm dg}=1/100$, dust grain ranges from 0.001 to 1 $\mu$m, and our fiducial dust mixture as described in section \ref{section:DustOpacitiesEddingtonRatios}. Only the fiducial model is presented here as other IMFs vary from these values by less than 10\%.}
\label{table:MeanOpacities}
\end{table}

\subsection{Eddington Ratios Spanning the optically-thin and single-scattering limits}
\label{section:EddRatioLimits}

\begin{figure*}
    \includegraphics[width = 0.47\textwidth]{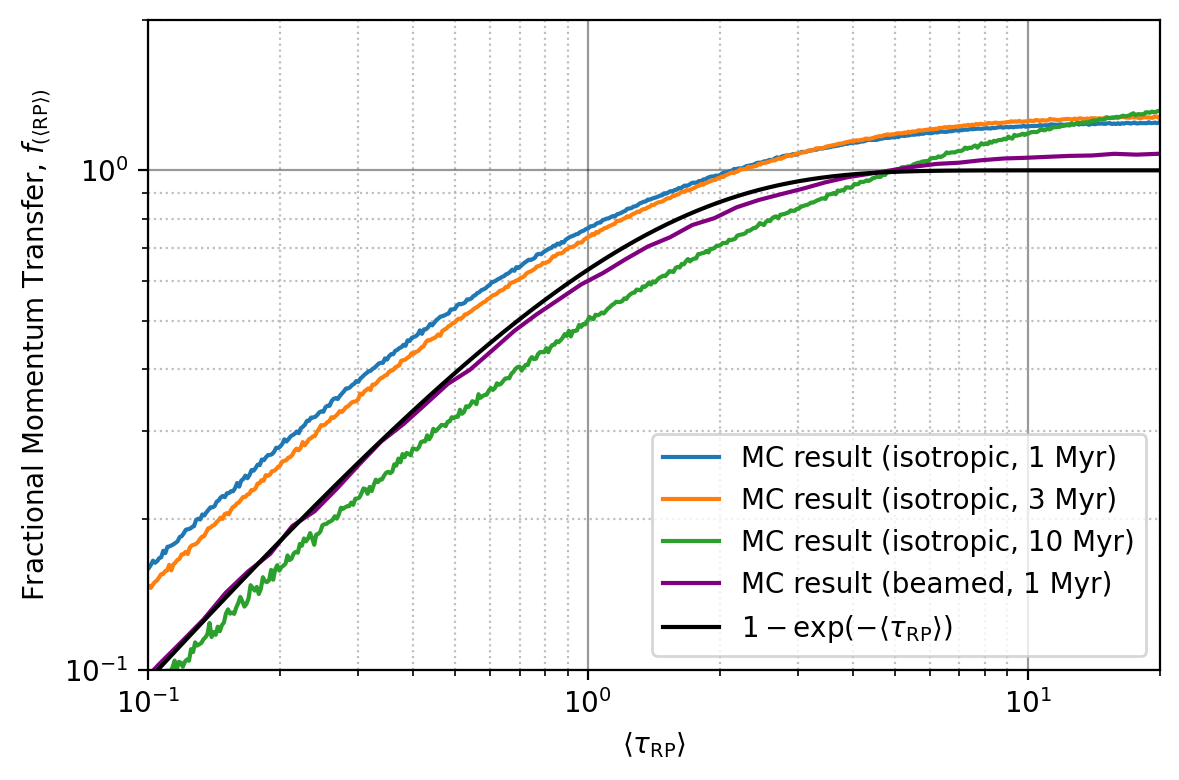}
    \includegraphics[width = 0.47\textwidth]{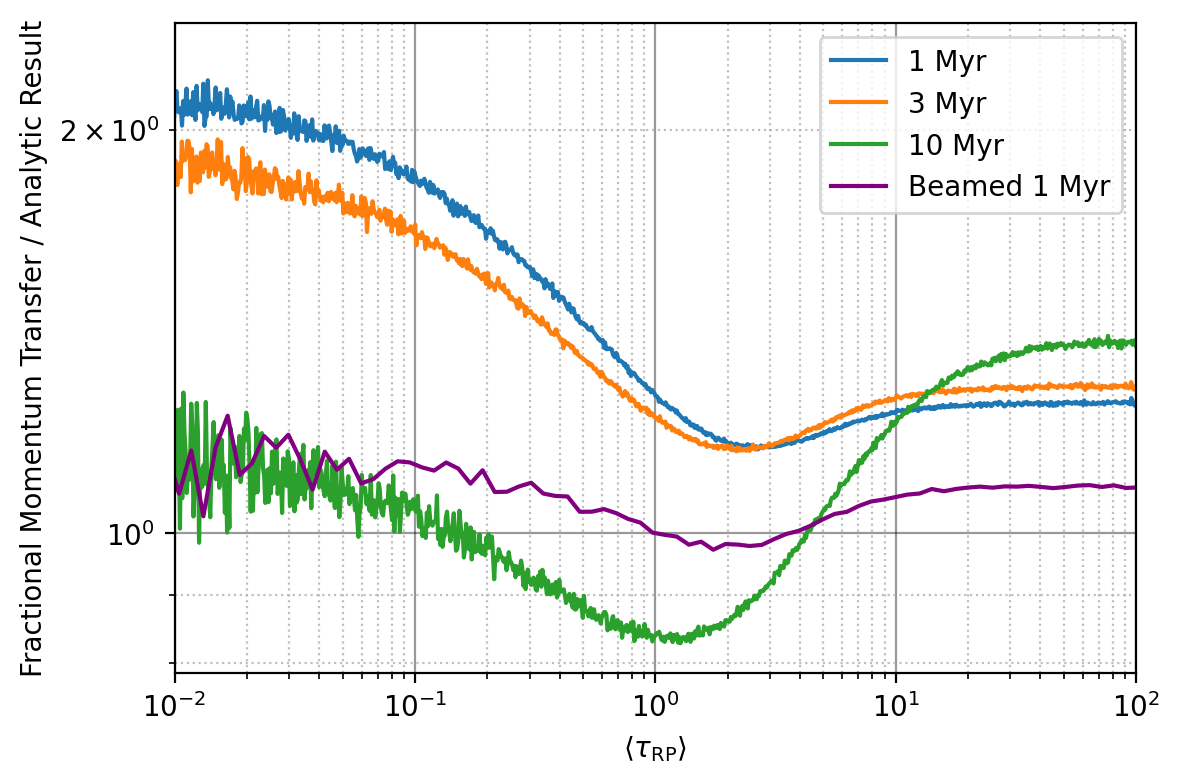}
    \caption{Left: Momentum transfer as a function of the mean atmospheric optical depth. Equation \ref{equation:AnalyticScattering} is used as an analytic approximation for the momentum transfer fraction, but these results show that for both optically thick and thin atmospheres we receive additional momentum deposition for isotropic emission. Right: The results of the Monte Carlo simulation divided by the analytic approximation in Equation \ref{equation:AnalyticScattering}. Each time is calculated by the BPASS SED for a stellar population of that age. The increase at thick atmospheres ($\tau > 1$) was expected from multiple scatterings. The increase at thin atmospheres was not expected, but comes from the same underlying physics. For both plots the data is for a stellar population using our fiducial IMF model. The calculation of $\langle\tau_{\rm RP}\rangle$ is dependent on the SED of the stellar model.}
    \label{figure:monte_carlo_result}
\end{figure*}

A commonly used analytic approximation for the fractional momentum transfer as a function of radiation pressure optical depth is  $f=1-\exp(-\tau)$, where $\tau = \kappa \Sigma_g$ and $\kappa$ is a constant opacity. As can be seen from Table \ref{table:MeanOpacities}, the radiation pressure flux-mean opacity is a function of time, and so a better approximation is
\begin{equation}
    f_{(\langle\tau_{\rm RP}\rangle)} = 1-e^{-\langle\tau_{\rm RP}\rangle},
    \label{equation:AnalyticScattering}
\end{equation}
where $\langle\tau_{\rm RP}\rangle$ depends explicitly on the stellar SED and is given in Equation \ref{equation:tau_av_RP}. In the optically-thin limit this reduces to $f_{(\langle\tau_{\rm RP}\rangle)}\simeq\langle\tau_{\rm RP}\rangle$, and in the optically-thick limit it becomes $f_{(\langle\tau_{\rm RP}\rangle)}\simeq1$. 

The goal of our Monte Carlo simulations is to test this analytic approximation, but including the additional physics of aging of the stellar population and multiple scattering. Figure \ref{figure:monte_carlo_result} presents the results of our simulations including full scattering, absorption, and spectral averaging, and compares with  Equation \ref{equation:AnalyticScattering}. The blue line presents the results for unbeamed, isotropic radiation at the injection boundary for a 1\,Myr old BPASS stellar population with our fiducial IMF (\S\ref{section:DustOpacitiesEddingtonRatios}, Table \ref{table:IMFModels}). The orange and green lines show this same stellar population aged to 3 and 10 Myr respectively. The purple line shows the 1 Myr fiducial model stellar population, but for beamed radiation, and the black line shows the approximation of Equation \ref{equation:AnalyticScattering}.

For the isotropic unbeamed case, we find that the fractional momentum transfer exceeds the beamed case in both the optically-thin and optically-thick limits. All optical depths given in this work are the optical depth for a vertically emitted photon. Photons emitted at shallow angles see a significantly thicker overall optical depth by $1/\cos(\theta)$ and this increased optical thickness of the vertical path through the atmosphere leads to a larger apparent optical depth for photons with shallow launch angles. Because the relation between the momentum transfer and optical depth plateaus at thick atmospheres, increasing the optical depth alone cannot explain the additional momentum found in simulations at thick atmospheres. By tracking the photon trajectories, we find that the excess momentum transfer comes from additional scatterings at the emission boundary. A photon is free to scatter between the upper and lower portions of the atmosphere, increasing momentum transfer efficiency, and photons released with $\cos\theta \simeq 0$ are more likely to scatter across this boundary multiple times enhancing the momentum transfer even at thin atmospheres.

\begin{figure}
    \includegraphics[width = 0.43\textwidth]{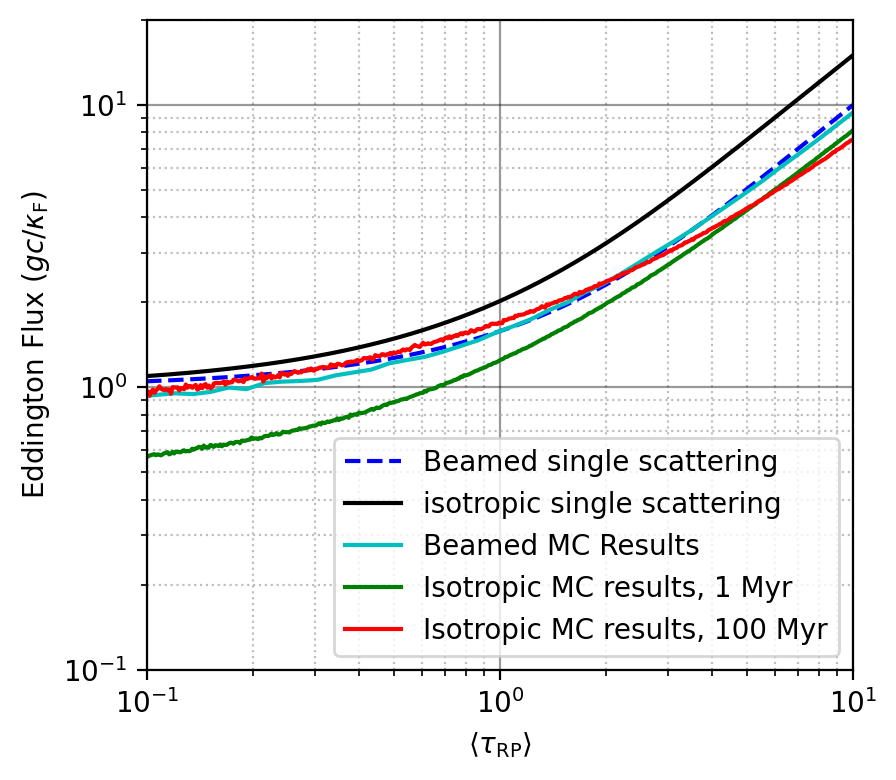}
    \caption{Shown here are the Eddington flux results of our Monte Carlo simulation for both a 1 Myr and 100 Myr fiducial population. Also shown for comparison are the beamed (dashed blue line) and isotropic (solid black line) single scattering results from \protect\cite{Wibking2018}. The simulation uses an isotropic source of photons, but more closely matches the beamed results due to a higher number of additional scatterings.}
    \label{figure:WibkingComparison}
\end{figure}

\subsection{Eddington Ratios}
\label{section:EddingtonRatios}

A general formula in planar geometry for the Eddington flux for direct radiation pressure in a column of dusty gas of height $H_{\rm gas}$ is
\begin{equation}
    F_{\rm Edd} = \frac{2\pi Gc\Sigma_{\rm gas}\Sigma_{\rm tot}}{f_{(\langle \rm RP \rangle)}},
    \label{equation:Eddington_flux_generic}
\end{equation}
where $f_{(\langle \rm RP \rangle)}$ is the momentum transfer efficiency calculated in the Monte Carlo simulation in plane-parallel geometry. This is a function of the spectrum averaged radiation pressure optical depth, $\langle\tau_{\rm RP}\rangle$. $\Sigma_{\rm tot}$ is the total surface mass density of gas and stars. \cite{Wibking2018} finds the similar equation for beamed radiation
\begin{equation}
    F_{\rm Edd} = \frac{gc}{\kappa_{\rm RP}}\frac{\langle\tau_{\rm RP}\rangle}{f_{(\langle\tau_{\rm RP}\rangle)}} = \frac{gc\langle\Sigma_{\rm gas}\rangle}{f_{(\langle\tau_{\rm RP}\rangle)}},
    \label{equation:Eddington_flux_Wibking}
\end{equation}
where $g = 2\pi G\Sigma_{\rm tot}$ is the gravitational acceleration from a thin uniform disk. \cite{Wibking2018} give this equation for the case of beamed radiation, rather than isotropic emission. Figure \ref{figure:WibkingComparison} shows the results from \cite{Wibking2018}, as well as the results from our fiducial Monte Carlo result for stellar populations at $1\,$Myr in cyan and green and $100\,$Myr in red. Our beamed results closely matches the beamed results from \cite{Wibking2018}. Our isotropic result for a young population, however, shows a reduction in the Eddington flux compared to \cite{Wibking2018}. This difference is due to additional scattering events contributing additional momentum to the atmosphere.

The momentum transfer efficiency depends ultimately on the surface mass density of the gas, since it is a strong function of $\langle\tau_{\rm RP}\rangle$. The results of the Monte Carlo simulation can be directly applied to galaxy data where these numbers can be measured or calculated.

\section{Application of Results to Galaxy Data}
\label{section:ApplicationToGalaxies}

\begin{figure*}
    \includegraphics[width=0.95\textwidth]{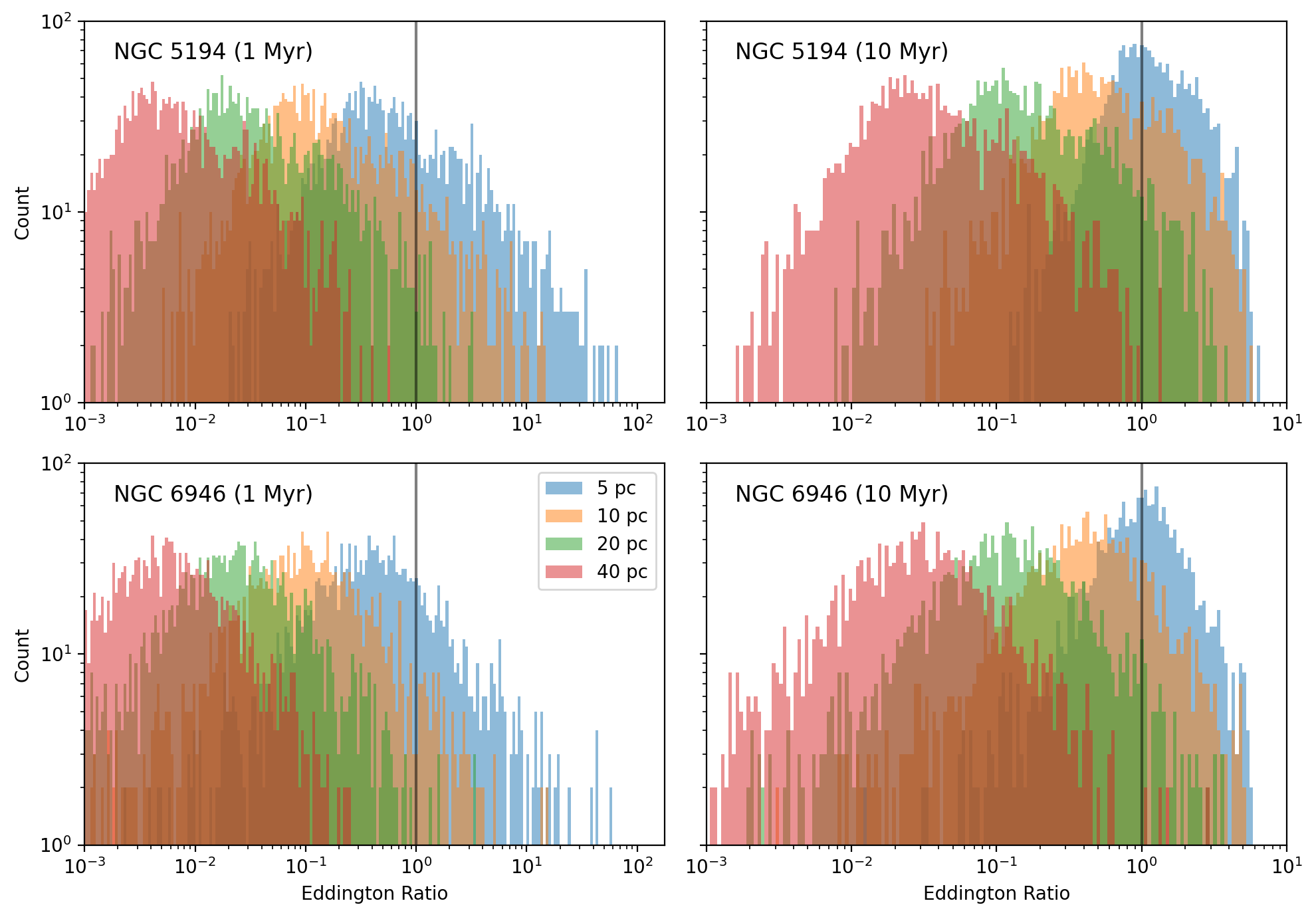}
    \caption{Histogram of the Eddington ratios of NGC 5194 and NGC 6946 for an assumption of spherical geometry and 4 different heights for the gas shell, $H_{\rm gas}$, across the galaxy and two different starting ages. Due to how the bolometric luminosity is calculated in Equation \ref{equation:L_bol} the luminosity increases as we assume an older starting age for the current observations because the ratio of H$\alpha$ to overall flux drops and the H$\alpha$ flux from measurements stays unchanged. Gas that is near the star cluster, with $H_{\rm gas}\approx5\,\textrm{pc}$ is more likely to be super-Eddington at either starting time. 40 pc is the approximate resolution of the measurements taken for each galaxy.}
    \label{figure:Edd_hist}
\end{figure*}

\begin{figure*}
    \includegraphics[width=0.95\textwidth]{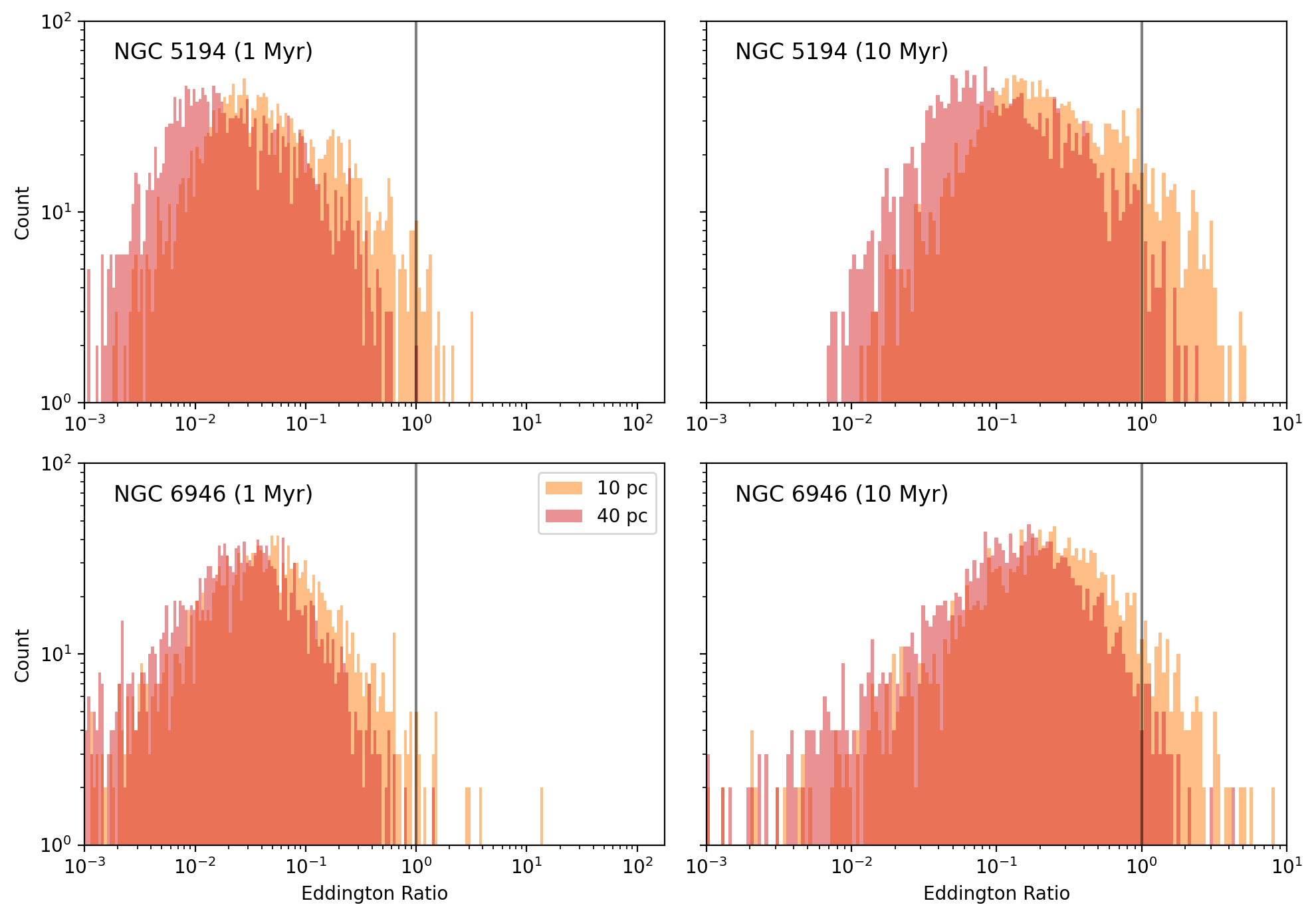}
    \caption{Histogram of the Eddington ratios of NGC 5194 and NGC 6946 for an assumption of planar geometry and 2 different heights for the gas shell, $H_{\rm gas}$, across the galaxy and two different starting ages. Since the bolometric flux (Equation \ref{equation:F_bol}) is calculated from the bolometric luminosity the same effect of increasing luminosity at 10 Myr can be seen here as in Figure \ref{figure:Edd_hist}.}
    \label{figure:Edd_hist_planar}
\end{figure*}

Here, we use the results from the previous subsections to evaluate the current Eddington ratios of galaxy sub-regions and then estimate the dynamics of dusty gas in those sub-regions. In doing so, we have to make assumptions about the geometrical distribution of gas and stars along the line of sight. Broadly, the geometries adopted can be thought of as (1) ``spherical," where we envision a single dominant star cluster powering each region and a dusty spherical expanding HII region or (2) ``planar," where we imagine the sources to be distributed in a geometrically thin sheet at the galaxy midplane. These two limiting geometries help highlight the uncertainties in assessing whether. or not sub-regions are super/sub-Eddington. In both the ``spherical" and ``planar" cases, we assume that the dusty gas probed by $A_{\rm H\alpha}$ along the line of sight is located a distance $H_{\rm gas}$ from the source of bolometric luminosity. In the spherical case, $H_{\rm gas}$ is the radial distance of a spherical shell of dusty gas from the central assumed source. In the planar case, $H_{\rm gas}$ is the vertical distance of a planar sheet above a uniform surface brightness region.

For simple assumptions, the ``spherical" picture predicts more total super-Eddington regions, with more rapid expansion for the regions that are super-Eddington, but with a strong dependence on the assumed value of $H_{\rm gas}$, the radial location of the dust along the line of sight. For the ``planar" model, we find less super-Eddington regions than the spherical model, and much slower dynamical expansion rates for those super-Eddington regions. The difference between these two limiting cases ultimately comes from how the flux is distributed spatially and how the total amount of old stellar mass enclosed by the dusty shell changes as a function of distance from the source of luminosity. We  treat each case separately below, but first discuss how observational quantities enter the calculations.

\subsection{Calculations from Observational quantities}
\label{section:ObservationalQuantities}

In order to constrain sub-region Eddington ratios from the \cite{Kessler2020} dataset, several physical quantities must be estimated. For each sub-region, a total H$\alpha$ luminosity $L_{\rm H\alpha,\, gal}$, extinction $A_{\rm H\alpha}$, and 3.6\,$\mu$m specific intensity are provided. The molecular gas surface density and HI surface density are also provided \citep{Walter2008,Leroy2009,Schuster2007,Schinnerer2013,Meyer2011,Rebolledo2015}. The apertures within which the H$\alpha$ data are measured correspond to radii for the sub-regions, $r_{\rm obs}$, of $40.4\,\textrm{pc}$ and $37.4\,\textrm{pc}$ at the assumed distances to NGC 5194 and NGC 6946. The 3.6\,$\mu$m Spitzer surface brightness, molecular gas surface density, and HI surface density measurements are all lower resolution than the H$\alpha$, and are taken to be the average value for the sub-regions, as in \cite{Kessler2020}. THINGS data for the HI measurements has $\sim450$\,pc resolution for NGC 5194 and $\sim200$\,pc resolution for NGC 6946. High-resolution 80\,pc CO data from PAWS is used for NGC 5194, and lower resolution 140\,pc CO data from CARMA is used for NGC 6946 \citep{Schuster2007,Walter2008,Leroy2009,Schinnerer2013,Rebolledo2015,Kessler2020}. These form our observational inputs and are used to estimate the bolometric luminosity of the driving region, the mass of new stars, the mass of the old stellar population, and the mass of the gas constituents, which include the dusty column projected along the line of sight and the molecular and atomic phases.

In this section, we treat each of the enclosed masses and derived quantities as if the geometry in each sub-region is ``spherical," where the mass of new stars that powers the H$\alpha$ in each sub-region is a central star cluster, and the obscuring dust is a spherical shell at some distance $H_{\rm gas}$ from that star cluster, embedded in a uniform distribution of old stars, CO, and HI, out to scale heights $H_{\rm old,\,\star}$, $H_{\rm CO}$, and $H_{\rm HI}$, respectively. As we show below, a critical uncertainty in the problem is the physical scale of the dusty column $H_{\rm gas}$, since this controls the amount of {\it old} stellar mass enclosed. In section \ref{section:planar_case}, we treat the same problem in a planar geometry to contrast with the spherical case presented here.

We scale the bolometric luminosity  $L_{\rm bol}$ of each sub-region to its H$\alpha$ luminosity using
\begin{equation}
    L_{\rm bol} = L_{\rm H\alpha,\, gal} \frac{L_{\rm bol,\, BPASS,(t)}}{L_{\rm H\alpha,\, BPASS,(t)}},
    \label{equation:L_bol}
\end{equation}
where $L_{\rm H\alpha,\, gal}$ is the measured H$\alpha$ luminosity of the sub-region. $L_{\rm bol,\, BPASS}$ and $L_{\rm H\alpha,\, BPASS}$ are the bolometric luminosity and H$\alpha$ luminosity of the BPASS model, respectively, at our selected age. For reference, observational constraints give this ratio as $L_{\rm bol}/L_{\rm H\alpha}\simeq0.00724^{-1}$ as an average across star-forming galaxies \citep{Kennicutt2012}, but for individual sub-regions $L_{\rm bol}/L_{\rm H\alpha}$ varies strongly as a function of age. For example, the BPASS models give  $L_{\rm bol}/L_{\rm H\alpha}\simeq0.0163^{-1}$, $0.0081^{-1}$, and $0.0021^{-1}$ at ages of 1\,Myr, 3\,Myr and 10\,Myr, respectively, for our fiducial IMF.\footnote{For the 135\_100 BPASS IMF these numbers are $0.0154^{-1}$, $0.0076^{-1}$, and $0.0021^{-1}$.} Because we do not have ages for the individual  sub-regions, we generally assume that they are $1-10$\,Myr old. Because we are starting with an observed value of $L_{\rm H\alpha}$, assuming the regions are older than $1-10$\,Myr results in very large inferred values of $L_{\rm bol}$.

A critical ingredient in evaluating the Eddington ratio of observed galaxy sub-regions is the old stellar mass enclosed by the dusty column along the line of sight to the sub-region. The observed $3.6$\,$\mu$m specific intensity is directly related to the {\it total} stellar mass column density along the line of sight by \citep{Leroy2019}
\begin{equation}
    \Sigma_{\rm old \, \star} = 350\frac{\rm M_\odot}{\rm pc^2} \left(\frac{I_{3.6\,\mu \rm m}}{\textrm{MJy}\,\textrm{sr}^{-1}}\right),
    \label{equation:Sigma_old_star}
\end{equation}
where $I_{\rm 3.6\,\mu m}$ is the observed specific intensity at $3.6\,\mu$m. We assume that the old stellar mass density is constant over a half height $H_{\rm old,\,\star}$ such that $\Sigma_{\rm old,\,\star}=2\rho_{\rm old,\,\star}H_{\rm old,\,\star}$. The total old stellar mass enclosed within a spherical region of radius $H_{\rm gas}$, the assumed location of the dusty column from the galaxy midplane, is then
\begin{equation}
    M_{\rm old,\,\star}(H_{\rm gas}) = \frac{2\pi}{3} \frac{\min(H_{\rm gas},\,H_{\rm old,\, \star})^3}{H_{\rm old,\, \star}}\Sigma_{\rm old \, \star}.
    \label{equation:M_old_star}
\end{equation}
\cite{kregel2002} give the scaleheight of the old stellar population as
\begin{equation}
    H_{\rm old,\,\star} = \frac{R_{\rm gal}}{7.3},
    \label{equation:H_old}
\end{equation}
where $R_{\rm gal}$ is the distance from the center of the galaxy to the sub-region. The old stellar mass enclosed by a dusty column in Equation \ref{equation:M_old_star} thus depends directly on two observables, $R_{\rm gal}$ and $\Sigma_{\rm old,\,\star}$, and an assumed value for the distance from the galaxy midplane to the dusty column, $H_{\rm gas}$.

There are additional contributions to the total enclosed mass that are related to the observables: the ``new" stellar mass associated with the source of the emission from the sub-region, the gas mass associated with the dusty column along the line of sight, and the mass of the cold gas column. The new stellar mass is directly related to the observed value of $L_{\rm H\alpha}$ and the assumed age of the underlying stellar population producing that emission through the BPASS models:
\begin{equation}
    M_{\rm new,\star} = L_{\rm bol}\frac{M_{\rm BPASS,(t)}}{L_{\rm bol,BPASS,(t)}} = L_{\rm H\alpha,\, gal} \frac{M_{\rm BPASS,(t)}}{L_{\rm H\alpha,\, BPASS,(t)}}.
    \label{equation:M_new_star}
\end{equation}
Note that in employing equations (\ref{equation:M_new_star}) and (\ref{equation:L_bol}, we assume that the IMF is fully-populated. This is an approximation that breaks down for $M_{\rm new,\,\star}\lesssim10^4$\,M$_\odot$. We return to a discussion of this issue in Section \ref{section:discussion}.

An estimate of the total gas mass in the shell associated with the projected dusty column can be made by assuming a spherical distribution
\begin{equation}
    M_{\rm gas} = \Sigma_{\rm gas}4\pi H_{\rm gas}^2,
    \label{equation:M_gas}
\end{equation}
where $\Sigma_{\rm gas}$ is given by Equation \ref{equation:Sigma_g}. $M_{\rm gas}$ depends strongly on the assumed value of $H_{\rm gas}$. Note that this approximation for the mass of gas assumes a geometrically-thin spherical shell. If we to instead imagine that the dusty absorbing gas uniformly fills a sphere of radius $H_{\rm gas}$ and that we see absorption against the central stellar population, we would obtain $M_{\rm gas} = \Sigma_{\rm gas}(4\pi/3) H_{\rm gas}^2$, a factor of 3 less than the approximation above. This factor can be important for low-mass regions, an issue we return to in Section \ref{section:discussion}.

Finally, the CO and HI gas mass can be calculated from measurements of the CO and HI surface brightness in each sub-region. These measurements are lower resolution than the H$\alpha$ measurements that define each sub-region. We use the same spatially-averaged surface densities used in \cite{Kessler2020}. We further assume for simplicity that the CO and HI gas is mixed on these scales and has a half height of $H_{\rm CO,\,HI}=100$\,pc (e.g., $\Sigma_{\rm CO,\,HI}=2\rho_{\rm CO,\,HI}H_{\rm CO,\,HI}$). More detailed models could be used with different CO and HI scale heights and different distributions as a function of galaxy radius. We discuss changes to this model in Section \ref{section:discussion}. The CO and HI masses enclosed by a spherical region of radius $H_{\rm gas}$ are then 
\begin{multline}
    M_{\rm CO,\,HI} 
    = \frac{2\pi}{3}\frac{\min\left(H_{\rm gas}, H_{\rm CO,HI} \,\rm pc\right)^3}{H_{\rm CO,HI}}\left(\Sigma_{\rm CO} + \Sigma_{\rm HI}\right).
    \label{equation:M_CO_HI}
\end{multline}

The final ingredient in calculating Eddington ratios for sub-regions is the value of $\langle \tau_{\rm RP}\rangle$, which is directly connected to the momentum transfer to the dusty column by our Monte Carlo calculations (Section \ref{section:MonteCarloResults}). The observed value of $A_{\rm H\alpha}$ can be converted directly into a dusty gas column density using Equation \ref{equation:Sigma_g}. When combined with the dust model, gives $\langle \tau_{\rm RP}\rangle$. The basic definition of $\langle\tau_{\rm RP}\rangle$ is given in Equation \ref{equation:tau_av_RP}, and from there we can construct a definition relating directly to $A_{\rm H\alpha}$ using
\begin{equation}
	\langle \tau_{\rm RP} \rangle_\lambda = f_{\rm dg}\frac{A_{\lambda}\int \pi a^2 \frac{dn}{da} \int (Q_{\rm abs, (\lambda)} + (1-g_{(a,\,\lambda)})Q_{\rm scatt,(\lambda)})L_{\lambda} d\lambda da}{1.086 L_{\rm bol} \int \pi a^2 \frac{dn}{da}\big(Q_{\rm abs,\, \lambda}+Q_{\rm scatt,\,\lambda}  \big)da}.
	\label{equation:tauRPFromObservales}
\end{equation}
With information about the underlying SED, and a dust model to give $Q_{\rm abs}$ and $Q_{\rm scatt}$ we can calculate the spectrum averaged optical thickness. In this paper we use $A_{\rm H\alpha}$ values from \cite{Kessler2020} to calculate our values of $\langle\tau_{\rm RP}\rangle$. In Equation \ref{equation:CriticalSigma} we define the critical gas column density for the ``single-scattering" limit. We can do the same for the extinction at a given wavelength using $A_\lambda = 1.086 \kappa_{\rm F,\,\lambda}\Sigma_{\rm gas}$ or
\begin{equation}
    A_{\lambda,\,(\langle\tau_{\rm RP}\rangle = 1)} \simeq 0.48\left(\frac{\kappa_{\rm F,\,\lambda}}{220\, {\rm cm^2\,\,g^{-1}}}\right)\left(\frac{500\,{\rm cm^2\,\,g^{-1}}}{\langle\kappa_{\rm RP}\rangle}\right).
    \label{equation:extinctionApproximation}
\end{equation}

\subsection{Eddington ratios and dynamics in the spherical case}
\label{section:spherical_case}

The Eddington luminosity for a spherical dusty shell is given by
\begin{equation}
    L_{\rm Edd}=\frac{G c M_{\rm tot}M_{\rm gas}}{f_{(\langle\tau_{\rm RP}\rangle)}}.
    \label{equation:Eddington_luminosity}
\end{equation}
Where $M_{\rm tot} = M_{\rm gas} + M_{\rm new\,,\star} + M_{\rm old\,,\star} + M_{\rm CO,\,HI}$, is the total mass enclosed inside the spherical shell, and $f_{(\langle\tau_{\rm RP}\rangle)}$ is the fraction of the photon momentum imparted to the dusty shell and is well-approximated by  $1-e^{-\langle\tau_{\rm RP}\rangle}$ (see Fig.~\ref{figure:monte_carlo_result}). In our calculations we replace this analytic result with our Monte Carlo results, $f_{(\langle \rm RP \rangle)}$ (see Section \ref{section:MonteCarloResults}).

\begin{figure*}
    \includegraphics[width = 0.95\textwidth]{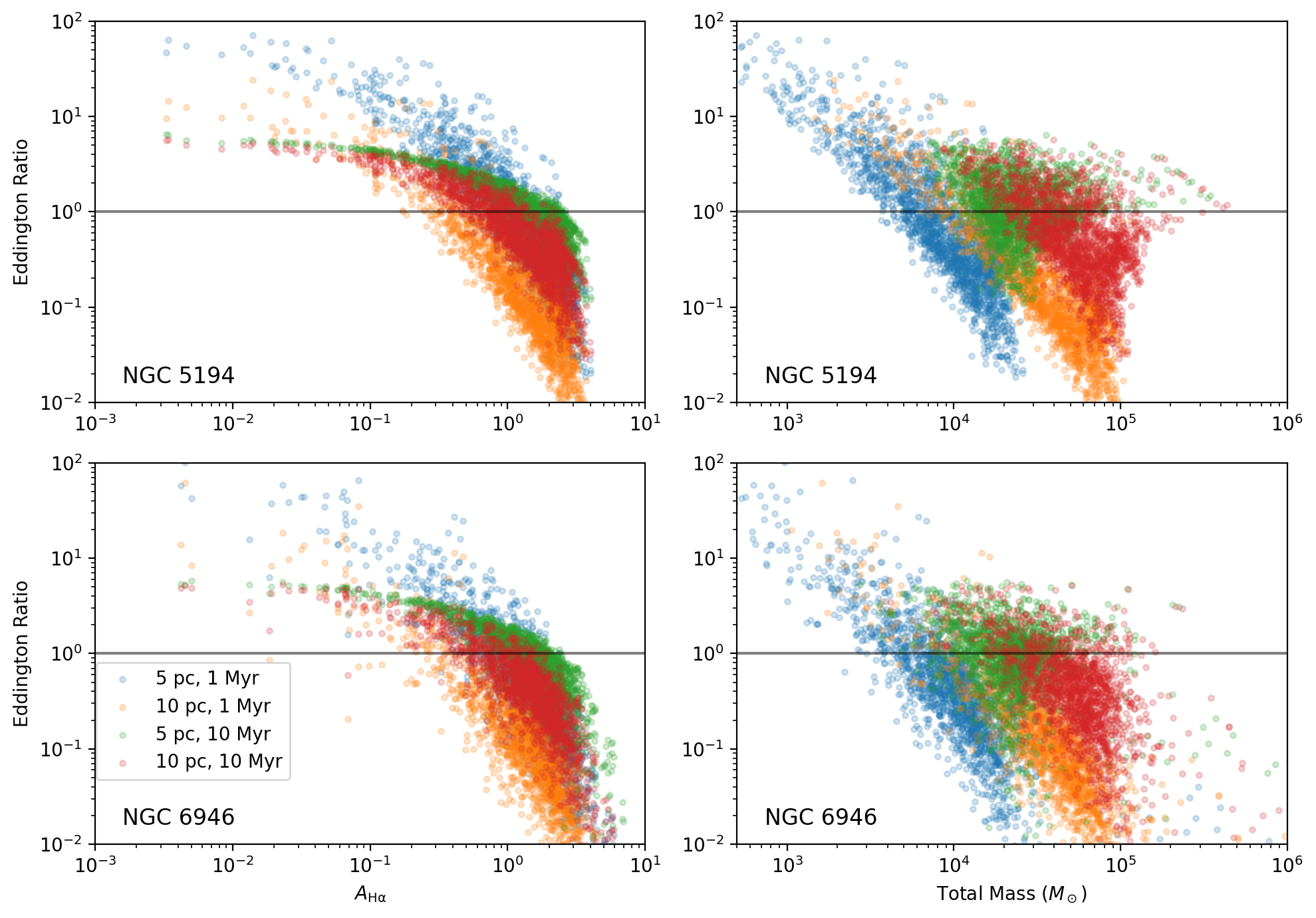}
    \caption{Eddington ratios for NGC 5194 and NGC 6946 as a function of the hydrogen-$\alpha$ extinction and total  mass enclosed by the gas shell. Shown are data points at 2 different ages, 1 and 10 Myr with starting gas radius of 5 and 10 pc. Similar to Figure \ref{figure:Edd_hist} there is an increase in the bolometric luminosity with age due to how the luminosity is calculated from measured data in Equation \ref{equation:L_bol}. Section \ref{section:ObservationalQuantities} has further discussion on this effect.}
    \label{figure:Edd_ratio_Aha}
\end{figure*}

To calculate the Eddington ratio $L_{\rm bol}/L_{\rm Edd}$, we take the size of the dusty shell $H_{\rm gas}$ to be a free parameter, and assume an age for the new stars to be 1 or 10\,Myr. We then compute $L_{\rm bol}$, the individual masses needed for $M_{\rm tot}$, and $f_{\langle\tau_{\rm RP}\rangle}$, as described in the previous sections. Figure \ref{figure:Edd_hist} shows the results for $L_{\rm bol}/L_{\rm Edd}$, for each of the sub-regions in NGC 5194 and NGC 6946, assuming spherical geometry, and for values of $H_{\rm gas}=5$, 10, 20, and 40\,pc (colors), and for assumed ages for the new stellar population of 1\,Myr (left) and 10\,Myr (right). 

The Eddington ratio distribution is strongly dependent on both $H_{\rm gas}$ and age. If the dusty gas is close to the host star cluster (small $H_{\rm gas}$) it is preferentially super-Eddington, predominantly because there is less old stellar mass enclosed within the region. For larger $H_{\rm gas}$, the mass budget is dominated by the old stellar population enclosed by $H_{\rm gas}$ and the region is sub-Eddington. If we imagine that the dusty column along the line of sight is not a thin shell, but is instead a radial distribution extending from close to the driving region to farther away, the gas close to the region will have large positive acceleration because it is super-Eddington (with a small amount of old stellar mass enclosed), while material further away will not be accelerated by radiation pressure because it is sub-Eddington. 

In almost all cases, the mass in old stars dominates the dynamics. In spherical geometry, the enclosed stellar mass scales as the cube of the radius of the gas distribution $H_{\rm gas}$, since the old stellar mass is assumed to be uniformly distributed across the scale height of the galactic disk $H_{\rm old,\star}$ (eq.~\ref{equation:H_old}).

Comparing the left and right panels of Figure \ref{figure:Edd_hist} shows the importance of the assumed age of the underlying stellar population. The assumed age directly affects the derived mass of new stars and the bolometric luminosity through the ratio $L_{\rm bol}/L_{\rm H\alpha}$. For $10\,$Myr, $L_{\rm bol}/L_{\rm H\alpha}$ is larger by a factor of 7.6 than $L_{\rm bol}/L_{\rm H\alpha}$ at $1\,$Myr. Thus, for a given observed $L_{\rm H\alpha}$, the bolometric luminosity and new stellar mass are both larger in the right hand panels for all regions than in the left hand panels. If we compare the $H_{\rm gas} = 5\,$pc distributions on the left and right, we see that the $10\,$Myr  distribution of Eddington ratios is narrower and more peaked, but that there are no regions with $L_{\rm bol}/L_{\rm Edd} \gtrsim10$. In contrast, for $H_{\rm gas} = 5\,$pc, the left panel contains a number of regions with $L_{\rm bol}/L_{\rm Edd} \gtrsim10$. This can be understood as a consequence of the change in $L_{\rm bol}/L_{\rm H\alpha}$ from 1 to 10 Myr as well. For $10\,$Myr, all regions have the luminosity and new stellar mass increased by the factor of 7.6 compared to $1\,$Myr, however in most regions $M_{\rm new\,\star}$ is a negligible contribution to the total mass. Regions with $L_{\rm bol}/L_{\rm Edd} \gtrsim10$ however, are regions where $M_{\rm new\,\star}$ is an appreciable factor in the total mass, leading to a decrease in $L_{\rm bol}/L_{\rm Edd}$ for those regions at $10\,$Myr compared to $1\,$Myr.

Note that the maximum assumed value of $H_{\rm gas}$ in Figure \ref{figure:Edd_hist} is $40\,$pc, which is approximately the size of the observed sub-regions. It is possible that the dusty gas is distributed on a larger vertical scale, e.g., $\sim 100-200$\,pc. In this case, the dusty gas would effectively ``see" the radiation from more than one individual sub-region. Multiple observed regions would contribute to the incident flux, but because the dusty gas has a larger physical scale, more mass is enclosed. 

We have investigated the effect of other regions on the Eddington ratios by summing their contributions, taking into account their distance to the selected region, assuming ages for the distant regions, and the effects of intervening dusty gas. Light from nearby clusters will begin to exert additional vertical force sooner than the light from distant clusters. When calculating the total luminosity of a region we can include a correction factor for the other regions in our data set as
\begin{equation}
     L_{\rm bol} = L_{\rm cl}\left(1-e^{-\langle\tau_{\rm RP}\rangle_{\rm cl}} + \sum_{\rm n}\frac{L_{\rm H\alpha,\, n}}{L_{\rm H\alpha,\, cl}}e^{-\langle\tau_{\rm RP}\rangle_{\rm n}}\sin\left(\arctan\left(\frac{H_{\rm gas}}{R_{\rm n}}\right)\right)\right).
     \label{equation:LuminosityOfAllClusters}
\end{equation}
Here we denote quantities from the cluster of interest with a cl subscript. We sum over each region scaling the luminosity of that region by the ratio of $A_{\rm H\alpha}$ for that region and the cluster we are interested in. We then multiply by a factor of $e^{\langle\tau_{\rm RP}\rangle}$ for each region to represent the optical thickness of that region, then a trigonometric factor to calculate the vertical component of the flux that reaches the region of interest. $R_{\rm n}$ is the distance from the reference cluster to the contributing cluster. We need to remove the over count for the cluster in question so we subtract a factor of $e^{\langle\tau_{\rm RP}\rangle_{\rm cl}}$. We then multiply luminosity of the cluster of interest.

It is assumed in this approach that the ratio of $L_{\rm H\alpha}$ for the regions will not change over time, however in practice this may not be the case. It would be possible to use the new stellar mass ratio, and account for varying SFR in each region but this was not done for this work. Additionally the optical depth of each regions should change over time as that region evolves, with the most luminous regions becoming more optically thin and allowing more light to reach the region of interest. This effect is not taken into account here, nor is the light from regions not included in the \cite{Kessler2020} data set making this a conservative estimate of the additional luminosity seen by the gas cloud. However, including the correction for the effects of additional regions does not change the conclusion that dusty gas with larger scale height will be substantially sub-Eddington because the dominant factor is the larger enclosed mass from old stars, and not the contribution to the incident flux from other regions.

We can also plot the derived Eddington ratios as a function of direct observational data. Figure \ref{figure:Edd_ratio_Aha} shows the Eddington ratio as a function of the H$\alpha$ extinction and total mass enclosed in the dusty shell at two different values of the shell radius and region age. In this figure we see that the Eddington ratio decreases as a function of both extinction and total mass. Because the surface density of gas is directly connected to $A_{\rm H\alpha}$, and because $F_{\rm Edd} \propto \Sigma_{\rm gas}/(1-\exp(-\langle\tau_{\rm RP}\rangle))$, it follows that all else equal, regions with larger $A_{\rm H\alpha}$ will have a lower Eddington ratio.

\begin{table*}
\centering
\begin{tabular}{ccccccc}
 Galaxy & Model & $L_{\rm bol}$ fraction & $M_{\rm gas}$ fraction & Model & $L_{\rm bol}$ fraction & $M_{\rm gas}$ fraction\\ 
   & Age, $H_{\rm gas}$ & super-Eddington & super-Eddington & Age, $H_{\rm gas}$ & super-Eddington & super-Eddington\\ 
 \hline\hline
\multirow{4}{*}{NGC 5194} & 1 Myr, 5 pc & 0.49093 & 0.14678 & 10 Myr, 5 pc & 0.68888 & 0.33829 \\
 & 1 Myr, 10 pc & 0.20391 & 0.02581 & 10 Myr, 10 pc & 0.38719 & 0.09414 \\
 & 1 Myr, 20 pc & 0.05262 & 0.00198 & 10 Myr, 20 pc & 0.12188 & 0.01103 \\
 & 1 Myr, 40 pc & 0.00000 & 0.00000 & 10 Myr, 40 pc & 0.01504 & 0.00030 \\ \hline
\multirow{4}{*}{NGC 6946} & 1 Myr, 5 pc & 0.36998 & 0.09712 &  10 Myr, 5 pc & 0.61328 & 0.27650 \\
 & 1 Myr, 10 pc & 0.12584 & 0.01277 & 10 Myr, 10 pc & 0.27628 & 0.05601 \\
 & 1 Myr, 20 pc & 0.03986 & 0.00088 & 10 Myr, 20 pc & 0.08341 & 0.00551\\
 & 1 Myr, 40 pc & 0.00594 & 0.00002 & 10 Myr, 40 pc & 0.02693 & 0.00041
\end{tabular}
\caption{Integrated quantities for NGC 5194 and NGC 6946 in spherical geometry at different assumed ages and column heights across the entire galaxy. The model columns denote the assumed age and column height of the gas shell in each region. A single age and height are taken for the entire galaxy for these ratios. These values are calculated by the sum of $L_{\rm bol}$ or $M_{\rm gas}$ in each super-Eddington sub-region and dividing by the sum of $L_{\rm bol}$ or $M_{\rm gas}$ in all sub-regions. The $L_{\rm bol}$ fraction is the fraction of the bolometric luminosity in all regions that comes from super-Eddington regions. $M_{\rm gas}$ fraction is the fraction of the total gas mass in super-Eddington regions.}
\label{table:integralQuantities}
\end{table*}

Table \ref{table:integralQuantities} shows the integrated quantities from all sub-regions in each galaxy in spherical geometry. These are the fraction of the luminosity or $M_{\rm gas}$ (Equation \ref{equation:M_gas}) across all sub-regions that resides in super-Eddington sub-regions. The fraction for $L_{\rm H\alpha}$ and $M_{\rm new}$ are identical to the fraction for $L_{\rm bol}$, as $L_{\rm bol}$ (Equation \ref{equation:L_bol}) and $M_{\rm new}$ (Equation \ref{equation:M_new_star}) are calculated from $L_{\rm H\alpha}$. As we assume a higher age for the sub-regions we find that more total regions are super-Eddington, which can be seen in Figure \ref{figure:Edd_hist}. As suggested by the discussion of Figure \ref{figure:Edd_hist}, assuming an age of $10$\,Myr increases the number of super-Eddington regions and this increase is proportional across all regions. This means that the increase in the fraction from 1 to 10 Myr is related only to the increase in the number of super-Eddington regions in those two models.

These results are only for the bright H$\alpha$ emitting regions included in our dataset. Eddington ratios for local star-forming galaxies like NGC 5194 and 6946 have been previously constructed by \cite{Andrews2011} and \cite{Wibking2018}. In those cases, the global flux was compared with the average surface density, including the molecular phase, which includes most of the mass. Yet, if regions are going to be super-Eddington, it is the low-column density sightlines that should be \citep{Thompson_Krumholz}. It is worth trying to connect what we have done in this paper to those previous estimates. For example, for NGC 5194 and 6946, \cite{Wibking2018} inferred $\Gamma_{\rm Edd}$ for the whole galaxy, with $\Sigma_{\rm gas}$ from molecular observations, $\Sigma_{\rm tot} = 10 \Sigma_{\rm gas}$, $L_{\rm bol} = L_{\rm FIR}$, and assuming the single-scattering limit. They found that $\Gamma_{\rm Edd} << 1$ for both galaxies. Here, we can take a different approach and ask about $\Gamma_{\rm Edd}$ by taking the whole galaxy luminosity in the optical and UV and using the globally inferred $E(B-V)$. With $\Gamma_{\rm Edd} = \kappa_{\rm UV+optical} F_{\rm UV+optical}/2\pi Gc\Sigma_{\rm tot}$. This gives $\Gamma_{\rm Edd} \sim 0.01$. This number can be compared with our median Eddington ratio from the individual sub-regions, where we find $0.54$, $0.13$, $0.03$ for $H_{\rm gas} = 5$, $10$, and $20\,$ pc. One uncertainty in this estimate is the fraction of $\Sigma_{\rm tot}$ that should be included since the old stellar scale height is larger than the young stellar scale and likely higher than scale of the dusty intervening medium.

These results can be contrasted with the planar geometry case, to which we now turn.

\subsection{Eddington ratios and dynamics in the planar case}
\label{section:planar_case}

It is instructive to construct Eddington ratios in an assumed planar-parallel geometry to contrast with the spherical calculations. In this picture, the mass of new stars that dominates the H$\alpha$ luminosity in each region is assumed to be geometrically thin and located at the galaxy midplane. The dusty column that produces the extinction $A_{\rm H\alpha}$ is located at a height $H_{\rm gas}$ above the midplane.  The dusty layer is embedded in a uniform distribution of old stellar mass that extends out to a height $H_{\rm old,\,\star}$, and a uniform distribution of cold/cool CO and HI gas that extends to a height of $H_{\rm CO,\,HI}$ above the midplane. We calculate the bolometric flux of a region,
\begin{equation}
    F_{\rm bol} = \frac{L_{\rm bol}}{\pi r_{\rm obs}^2},
    \label{equation:F_bol}
\end{equation}
where the bolometric luminosity is given by Equation \ref{equation:L_bol}, and $r_{\rm obs}$ is the radius of the sub-region, which we take to be $r_{\rm obs}=40$\,pc since the nominal values for NGC 5194 and NGC 6946 are $\simeq40.4$\,pc for  and $\simeq37.4$\,pc, respectively.

Starting from Equation \ref{equation:Eddington_flux_generic} we can expand the total surface mass density and use observational data to calculate its constituent parts. The Eddington flux in each region is then
\begin{equation}
    F_{\rm Edd}(H_{\rm gas}) = \frac{2\pi Gc\Sigma_{\rm gas}\Sigma_{\rm tot}(H_{\rm gas})}{f_{(\langle\tau_{\rm RP}\rangle)}},
    \label{equation:F_Edd}
\end{equation}
where 
\begin{eqnarray}
    \Sigma_{\rm tot}(H_{\rm gas})&=&\Sigma_{\rm old,\star}\min\left(1,\frac{H_{\rm gas}}{H_{\rm old\,,\star}}\right) + \Sigma_{\rm CO,\,HI}\min\left(1,\frac{H_{\rm gas}}{H_{\rm CO,HI}}\right)   \nonumber \\   &\,& +\,\Sigma_{\rm new,\star} + \Sigma_{\rm gas}.
\end{eqnarray}
Here, $\Sigma_{\rm old,\star}$ and $\Sigma_{\rm new,\star}$ are the ``old" and ``new" stellar mass surface densities, and $\Sigma_{\rm CO,\,HI}$ is the cold gas surface density. $H_{\rm CO,HI}=100$\,pc is the assumed scale height of the cold CO and HI gas. The terms $\textrm{min}(1,\frac{H_{\rm gas}}{H_{\rm old,\star}})$ and $\textrm{min}(1,\frac{H_{\rm gas}}{H_{\rm CO,HI}})$ represent the gas encompassing more old stellar mass and cold gas for an assumed value of the height of the gas shell, $H_{\rm gas}$, but once it reaches the maximum height of the respective mass column the surface density below it stays constant. $\Sigma_{\rm old,\star}$ is given by equation (\ref{equation:Sigma_old_star}). $\Sigma_{\rm old,\star}$ is directly related to the 3.6$\mu$m surface brightness in each sub-region by equation (\ref{equation:Sigma_old_star}). $\Sigma_{\rm new,\star}$ is directly related to the surface brightness of H$\alpha$ within each aperture:
\begin{equation}
    \Sigma_{\rm new,\star} = \frac{M_{\rm new,\star}}{\pi r_{\rm obs}^2},
    \label{equation:Sigma_new_star}
\end{equation}
where $M_{\rm new,\star}$ is given in equation (\ref{equation:M_new_star}). As in the spherical case, the column density of CO and HI are taken from \cite{Kessler2020}. We note again that both the CO and HI data are lower-resolution than the H$\alpha$ data (Section \ref{section:ObservationalQuantities}).

As in the spherical case, when calculating the Eddington flux ratios, $F_{\rm bol}/F_{\rm Edd}$, $H_{\rm gas}$ is a free parameter. Figure \ref{figure:Edd_hist_planar} shows the Eddington flux ratio for each sub-region in NGC 5194 and NGC 6946, assuming planar geometry, for values of $H_{\rm gas} = 10$ and $40$\,pc and stellar population ages of $1$\,Myr (left) and $10$\,Myr (right). We see that assuming an older population increases the Eddington ratios of all regions. This shift is caused by the increase in the bolometric luminosity from the $L_{\rm bol}/L_{\rm H\alpha}$ ratio changing by a factor of 7.6. Unlike the spherical case the planar model has few highly super-Eddington regions, but is less sensitive to changes in the assumed column height. Comparing Figures \ref{figure:Edd_hist} and \ref{figure:Edd_hist_planar} we see that the in the spherical case there are few if any super-Eddington regions assuming the maximum model size of $40\,$pc. This effect is from the reduced dependence of the planar equations on the height of the gas shell.

Table \ref{table:integralQuantitiesPlanar} shows the integrated quantities for sub-regions in planar geometry. These are the fraction of the flux or $M_{\rm gas}$ (Equation \ref{equation:M_gas}) across all sub-regions that resides in super-Eddington sub-regions. The fraction for $F_{\rm H\alpha}$ and $M_{\rm new}$ are identical to the fraction for $F_{\rm bol}$, as $F_{\rm H\alpha}$, $F_{\rm bol}$ (Equation \ref{equation:F_bol}), and $M_{\rm new}$ (Equation \ref{equation:M_new_star}) are calculated from $L_{\rm H\alpha}$. As in the discussion of Table \ref{table:integralQuantities} there is an increase in the number of super-Eddington regions at $10\,$Myr compared to $1\,$Myr due to an increase in $L_{\rm bol}/L_{\rm H\alpha}$ in the $10\,$Myr model from the $1\,$Myr model. We can see that for the $40$\,pc model $0.6\%$ of the total flux in all regions of NGC 5194 belongs to super-Eddington regions assuming an age of 1 Myr, and this changes to $6\%$ assuming an age of 10 Myr. For a column height of $100\,$pc we do not see any super-Eddington regions in either model. This calculation is complicated by the effects of neighboring regions becoming relevant, as discussed in Sections \ref{section:spherical_case} and \ref{section:modeling_cloud_velocities}.

\begin{table*}
\centering
\begin{tabular}{ccccccc}
 Galaxy & Model & $F_{\rm bol}$ fraction & $M_{\rm gas}$ fraction & Model & $F_{\rm bol}$ fraction & $M_{\rm gas}$ fraction\\ 
   & Age, $H_{\rm gas}$ & super-Eddington & super-Eddington & Age, $H_{\rm gas}$ & super-Eddington & super-Eddington\\ 
 \hline\hline
\multirow{4}{*}{NGC 5194} & 1 Myr, 5 pc & 0.08815 & 0.00539 & 10 Myr, 5 pc & 0.25624 & 0.04219 \\
 & 1 Myr, 10 pc & 0.06172 & 0.00299 & 10 Myr, 10 pc & 0.18222 & 0.02383 \\
 & 1 Myr, 20 pc & 0.03043 & 0.00103 & 10 Myr, 20 pc & 0.12174 & 0.01281 \\
 & 1 Myr, 40 pc & 0.00609 & 0.00014 & 10 Myr, 40 pc & 0.06654 & 0.00418 \\ \hline
\multirow{4}{*}{NGC 6946} & 1 Myr, 5 pc & 0.06484 & 0.00320 &  10 Myr, 5 pc & 0.20678 & 0.03079 \\
 & 1 Myr, 10 pc & 0.04853 & 0.00173 & 10 Myr, 10 pc & 0.17836 & 0.02368 \\
 & 1 Myr, 20 pc & 0.03787 & 0.00099 & 10 Myr, 20 pc & 0.13891 & 0.01489\\
 & 1 Myr, 40 pc & 0.03122 & 0.00065 & 10 Myr, 40 pc & 0.09038 & 0.00694
\end{tabular}
\caption{Integrated quantities for NGC 5194 and NGC 6946 in planar geometry at different assumed ages and column heights across the entire galaxy. The model columns denote the assumed age and column height of the gas shell in each region. A single age and height are taken for the entire galaxy for these ratios. These values are calculated by the sum of $F_{\rm bol}$ or $M_{\rm gas}$ in each super-Eddington sub-region and dividing by the sum of $F_{\rm bol}$ or $M_{\rm gas}$ in all sub-regions. The $F_{\rm bol}$ fraction is the fraction of the bolometric luminosity in all regions that comes from super-Eddington regions. $M_{\rm gas}$ fraction is the fraction of the total gas mass in super-Eddington regions.}
\label{table:integralQuantitiesPlanar}
\end{table*}

\label{section:additional_pressures}
 
 \subsection{Modeling Cloud Velocities}
\label{section:modeling_cloud_velocities}

\begin{figure*} 
    \includegraphics[width=0.95\textwidth]{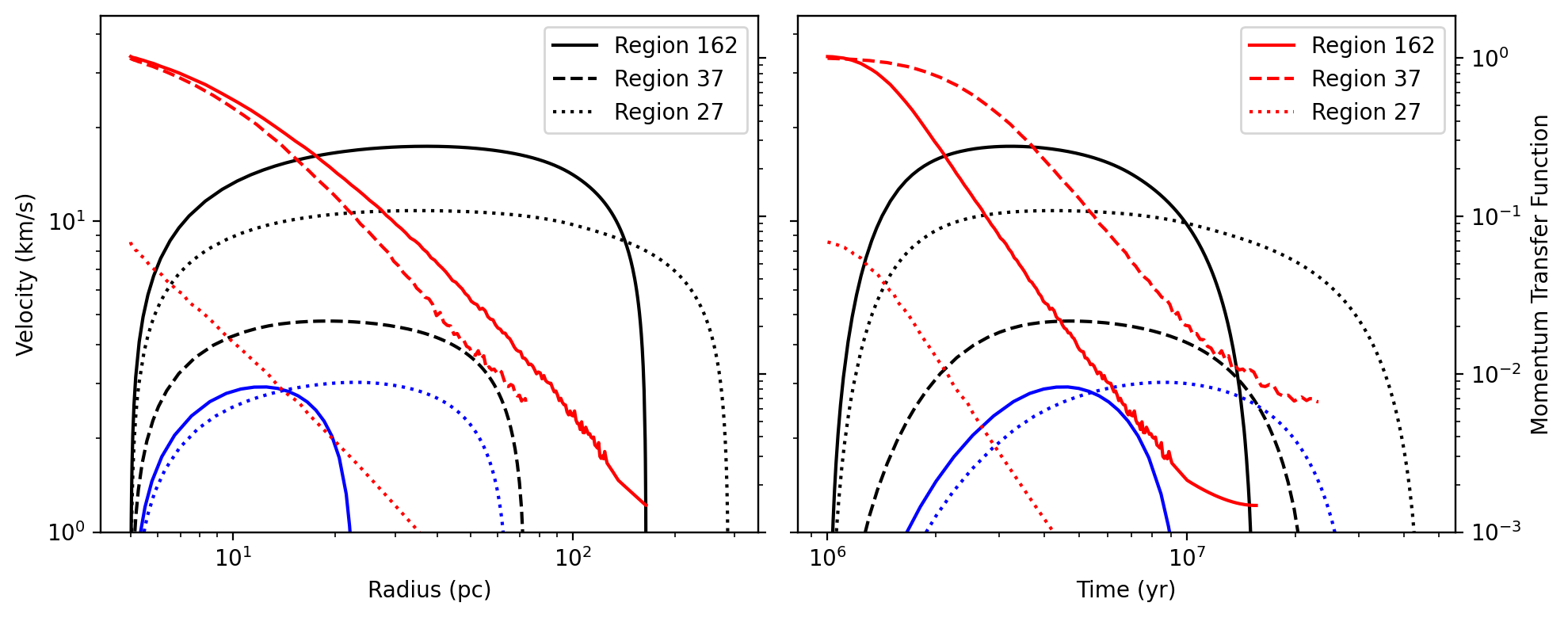}
    \caption{Velocity as a function of radius and time for regions 37 and 162 in NGC 5194 and region 27 in NGC 6946. Gas is assumed to start at 5 pc with a 1 Myr stellar population. All three regions are super Eddington using these conditions, and the details for these regions are given in Table \ref{table:GalaxyRegions}. The velocities shown in this figure are for an aging stellar population. As populations age and redden the radiation pressure decreases, however in the spherical case shown in the black lines, this aging has little effect on the overall velocity or final radius as the initial acceleration is high. In the planar model shown in the blue lines the gas fails to reach a high velocity before stellar aging causes the region to become sub Eddington, preventing long term outward movement. The momentum transfer function is shown in red. This is the result of the Monte Carlo simulation for the region's SED and optical thickness.}
    \label{figure:velocity_plot}
\end{figure*}

\begin{table*}
\centering
\begin{tabular}{cccccccccc}
Galaxy & Region & $H_{\rm gas}$\,(pc) & $L_{\rm bol}\, (L_\odot)$ & $\rm A_{\rm H\alpha}$ & $M_{\rm gas}\, (M_\odot)$ & $M_{\rm new\,\star}\, (M_\odot)$ & $M_{\rm old\,\star}\, (M_\odot)$ & Eddington Ratio & $v_{\rm max}$ (km/s) \\ \hline \hline
\multirow{8}{*}{NGC 5194} & \multirow{4}{*}{37} & 5 & \multirow{4}{*}{$5\times 10^{5}$} & \multirow{4}{*}{0.25} & 1500 & \multirow{4}{*}{380} & 90 & 11 & 4.7 \\
 &  & 10 &  &  & 6200 &  & 720 & 3.0 & 1.1 \\
 &  & 20 &  &  & $2.4\times 10^4$ &  & 5700 & 0.6 & - \\
 &  & 40 &  &  & $9.8\times 10^4$ &  & $4.6\times 10^4$ & 0.1 & - \\ \cline{2-10} 
 & \multirow{4}{*}{162} & 5 & \multirow{4}{*}{$4.2\times 10^{6}$} & \multirow{4}{*}{0.27} & 1700 & \multirow{4}{*}{2900} & 290 & 34 & 17 \\
 &  & 10 &  &  & 6700 &  & 2300 & 13 & 8.8 \\
 &  & 20 &  &  & $2.7\times 10^4$ &  & $18\times 10^3$ & 3.1 & 2.0 \\
 &  & 40 &  &  & $1.1\times 10^5$ &  & $1.4\times 10^5$ & 0.5 & - \\ \hline
\multirow{4}{*}{NGC 6946} & \multirow{4}{*}{27} & 5 & \multirow{4}{*}{$1.2\times 10^{6}$} & \multirow{4}{*}{$4.5\times 10^{-3}$} & 28 & \multirow{4}{*}{860} & 28 & 100 & 11 \\
 &  & 10 &  &  & 110 &  & 230 & 61 & 5.6 \\
 &  & 20 &  &  & 450 &  & 1800 & 15 & 1.9 \\
 &  & 40 &  &  & 1800 &  & $1.4\times 10^4$ & 2.2 & 0.2
\end{tabular}
\caption{Calculated and measured properties of selected high velocity galaxy regions at several starting gas heights. Regions are assumed to be young, 1 Myr, stellar populations. Table \ref{table:GalaxyRegions10Myr} shows this same data for an assumed age of 10 Myr.}
\label{table:GalaxyRegions}
\end{table*}

\begin{table*}
\centering
\begin{tabular}{cccccccccc}
Galaxy & Region & $H_{\rm gas}$\,(pc) & $L_{\rm bol}\, (L_\odot)$ & $\rm A_{\rm H\alpha}$ & $M_{\rm gas}\, (M_\odot)$ & $M_{\rm new\,\star}\, (M_\odot)$ & $M_{\rm old\,\star}\, (M_\odot)$ & Eddington Ratio & $v_{\rm max}$ (km/s) \\ \hline \hline
\multirow{8}{*}{NGC 5194} & \multirow{4}{*}{37} & 5 & \multirow{4}{*}{$4.1\times 10^{6}$} & \multirow{4}{*}{0.25} & 1500 & \multirow{4}{*}{$2.2\times 10^4$} & 90 & 3.6 & 9.6 \\
 &  & 10 &  &  & 6200 &  & 720 & 2.9 & 4.7 \\
 &  & 20 &  &  & $2.4\times 10^4$ &  & 5800 & 1.5 & 0.4 \\
 &  & 40 &  &  & $9.8\times 10^4$ &  & $4.6\times 10^4$ & 0.4 & - \\ \cline{2-10} 
 & \multirow{4}{*}{162} & 5 & \multirow{4}{*}{$3.2\times 10^7$} & \multirow{4}{*}{0.27} & 1700 & \multirow{4}{*}{$1.7\times 10^5$} & 290 & 3.7 & 30 \\
 &  & 10 &  &  & 6800 &  & 2300 & 3.6 & 18 \\
 &  & 20 &  &  & $2.7\times 10^4$ &  & $1.8\times 10^4$ & 2.9 & 8.1 \\
 &  & 40 &  &  & $1\times 10^5$ &  & $1.4\times 10^5$ & 1.3 & 0.4 \\ \hline
\multirow{4}{*}{NGC 6946} & \multirow{4}{*}{27} & 5 & \multirow{4}{*}{$9.4\times 10^{6}$} & \multirow{4}{*}{$4.5\times 10^{-3}$} & 28 & \multirow{4}{*}{$5\times 10^4$} & 28 & 5.2 & 18 \\
 &  & 10 &  &  & 110 &  & 230 & 5.2 & 11 \\
 &  & 20 &  &  & 450 &  & 1800 & 4.7 & 5.5 \\
 &  & 40 &  &  & 1800 &  & $1.4\times 10^4$ & 2.8 & 1.5
\end{tabular}
\caption{Calculated and measured properties of selected high velocity galaxy regions at several starting gas heights for a starting age of 10 Myr. Table \ref{table:GalaxyRegions} shows this same data for an assumed age of 1 Myr.}
\label{table:GalaxyRegions10Myr}
\end{table*}

For any of the super-Eddington regions we can model the dynamics of the intervening dusty gas as it is accelerated (e.g., \citealt{Murray2011,Thompson2015}). In the spherical case, the momentum equation for a cold dusty shell of mass $M_{\rm gas,\,0}$ is given by
\begin{equation}
    v\frac{dv}{dr}= -\frac{G M_{\rm tot}(<r)}{r^2} + \frac{f_{(\langle \tau_{\rm RP} \rangle)}L_{\rm bol}}{cM_{\rm gas,\,0}},
    \label{equation:velocity_eq_spherical}
\end{equation}
In the above, $dt = dr/v$, $r$ is the shell radius, $H_{\rm gas,0}$ is the initial shell radius at  $t = t_0$,  $f_{(\langle \tau_{\rm RP} \rangle)}$ is the Monte Carlo result for that column density, and the total dynamical mass enclosed at radius $r$ is given by 
\begin{multline}
    M_{\rm tot}(<r) = M_{\rm old, \star}\frac{\min(H_{\rm old,\star},r)^3}{H_{\rm gas,0}^3} + M_{\rm CO,HI}\frac{\min(H_{\rm CO,HI},r)^3}{H_{\rm gas,0}^3} \\
    + M_{\rm gas,\,0} + M_{\rm new, \star},
\end{multline}
Where $M_{\rm old, \star}$ is the initially enclosed old stellar mass given by Equation \ref{equation:M_old_star} and $M_{\rm CO,HI}$ is the initially enclosed cold gas given by Equation \ref{equation:M_CO_HI}.

As the shell moves outward, its column density and optical depth decrease, and the star cluster providing the radiation pressure age, changing their bolometric luminosity and SED shape. The luminosity and SED changes are easily incorporated by interpolating the BPASS models as a function of time as the shell evolves. The Monte Carlo results for the momentum coupling are more complicated because the column density decreases as the shell expands and the radiation pressure opacity changes as the stellar population ages. Both of these effects are accounted for in the spherical geometry by continuously updating the column density and radiation pressure mean opacity as the shell expands.

As in our discussion of the Eddington ratio distributions in Sections \ref{section:spherical_case} and \ref{section:planar_case}, it is instructive to compare Equation \ref{equation:velocity_eq_spherical} in the spherical approximation with the dynamics for super-Eddington regions assuming a planar geometry. The equation of motion for a planar sheet of dusty gas accelerated by a planar distribution of sources with flux $F_{\rm bol}$ is
\begin{equation}
    v\frac{dv}{dr} = - 2\pi G\Sigma_{\rm tot} + \frac{f_{(\langle \rm RP \rangle)}F_{\rm bol}}{c\Sigma_{\rm g}},
    \label{equation:v_dvdr}
\end{equation}
with $F_{\rm bol}$ being the bolometric flux from Equation \ref{equation:F_bol}, the bolometric luminosity per area. There are two critical differences here between the spherical and planar geometry. The first is that, from the observations, we would take the flux to be spread uniformly over the observed subregion, which is approximately 40\,pc in radius. This is qualitatively different from the spherical case, where the luminosity is assumed to come from a compact central source and where we give ourselves the freedom to imagine a spherical shell expanding from scales smaller than the radius of the subregion (e.g., 5, 10, or 20\,pc). The acceleration term is correspondingly much smaller for each subregion and the expected asymptotic velocity is much smaller than in the spherical case. Second, unlike the spherical case, in the planar picture, the column density of the projected gas is constant as it is accelerated upward. 

The results of these velocity calculations for 3 highly super-Eddington regions can be seen in Figure \ref{figure:velocity_plot}. The black lines show the velocity in the spherical model, the blue lines are the velocity in the planar model, and the red lines are the spherical model's associated momentum transfer functions, $f_{\rm \langle RP\rangle}$ using our Monte Carlo results. The line for the planar model for region 37 does not appear because the dusty shell's velocity never exceeds $1\,$km/s. The left panel shows these velocities as a function of distance from the central stellar cluster (in the spherical model) or the galaxy midplane (in the planar model), while the right panel shows the velocity as a function of time. In both the spherical and planar models the acceleration of material takes place over a a few pc before acceleration stops. All regions in both models come to a stop by $40\,$Myr, and achieve their maximum velocity by $10\,$Myr. The effects of stellar aging on the luminosity and SED begin after $3\,$Myr. The stellar aging, combined with the fact that the shell encompasses more old stellar mass and cold gas as the shell expands outward prevents all but a few regions with high Eddington ratios from expanding past the $40\,$pc scale of the observed sub-region size.

For those that do expand past $40$\,pc scales, we also investigate the effect of other regions using Equation \ref{equation:LuminosityOfAllClusters}. Because we are accounting for the additional luminosity of these regions we also need to account for the additional mass enclosed as the shell expands to larger scale.  We can modify Equation \ref{equation:v_dvdr} to include a term for the effects of the mass from additional neighboring and distant regions, similar to the construction of Equation \ref{equation:LuminosityOfAllClusters} we find that 
\begin{multline}
    v\frac{dv}{dr}= -\frac{G M_{\rm tot}(<r)}{r^2} + \frac{f_{(\langle \tau_{\rm RP} \rangle)}L_{\rm bol}}{cM_{\rm gas,\,0}}  \\ - G\sum_{n \neq cl}\frac{M_{\rm tot, n}}{H_{\rm gas}^2 + R_n^2}\sin\left(\tan^{-1}\left(\frac{H_{\rm gas}}{R_{\rm n}}\right)\right),
    \label{equation:velocity_eq_spherical_whole_galaxy}
\end{multline}
where we sum over all regions but the cluster of interest, as its mass is accounted for in the first term already. We checked the contribution from additional galaxy regions and found that it does not qualitatively change the results.

\section{Discussion and Conclusion}
\label{section:discussion}

In this paper we attempt to evaluate the dynamical importance of radiation pressure on dust in thousands of H$\alpha$-emitting sub-regions across two local star-forming galaxies, NGC 6946 and NGC 5194.

Using wavelength-dependent anisotropic scattering Monte Carlo calculations we show that $1-\exp(-\langle\tau_{\rm RP}\rangle)$ is a good approximation to the fraction of the radiation momentum deposited in a dusty column, ranging from optically-thin through the single-scattering limit (see Figure \ref{figure:monte_carlo_result}). We compute values of $\langle\kappa_{\rm RP}\rangle$ for SEDs of specified age from BPASS in Table \ref{table:MeanOpacities} (see also Appendix \hyperref[appendix:DustGrains]{A}). As shown in right-hand panel of Figure \ref{figure:KappaAndLOverM}, for standard IMFs, optically-thin sightlines to newly-formed star clusters are $\simeq10-50$ times super-Eddington for $\sim4$\,Myr after birth and remain super-Eddington for $\simeq30$\,Myr. For continuous star formation optically-thin sightlines remain super-Eddington for $\simeq 200$\, Myr, assuming no mass is lost from the system considered.

For the individual sub-regions in NGC 5194 and 6946, we use the observed values of $L_{\rm H\alpha}$, the extinction $A_{\rm H\alpha}$, the total stellar surface density projected for each region, and the average local projected HI and molecular gas surface densities to calculate the Eddington ratio  (data from \citealt{Kessler2020}). Figures \ref{figure:Edd_hist} and \ref{figure:Edd_hist_planar} show the distribution of Eddington ratios in sub-regions for different values of the age of the sub-regions and the height of the dusty shell. Tables \ref{table:integralQuantities} and \ref{table:integralQuantitiesPlanar} show the fraction of the bolometric luminosity and masses in all studied sub-regions that are super-Eddington. From these we can see that the underlying geometry, choice of age for the stellar cluster, and the height of the dusty shell has a large impact on whether a region will be super-Eddington. For the largest assumed values for the height of the projected dusty gas ($40$\,pc), $\simeq0-3$\% of sightlines are super-Eddington, but for an assumed height two times smaller (20\,pc), $4-12$\% of sightlines are super-Eddington.

This picture of the potential importance of dust radiation pressure contrasts sharply with galaxy-averaged estimates (e.g., \citealt{Andrews2011,Wibking2018}), which would indicate that such systems are very sub-Eddington. For example, using numbers from \cite{Kennicutt1998} for the surface density of star formation and the surface density of gas for NCG 5194, we estimate a galaxy-averaged single-scattering Eddington flux of $F_{\rm Edd}\simeq2\pi G\Sigma_{\rm tot}\Sigma_{\rm gas}c\simeq1\times10^{10}\,{\rm L_\odot/kpc^2}(\Sigma_{\rm gas}/30\,{\rm M_\odot/pc^2})^2(0.1/f_{\rm gas})$, where we have assumed that the total surface density is related to the gas surface density by $\Sigma_{\rm tot}=\Sigma_{\rm gas}/f_{\rm gas}$. This Eddington flux is a factor of $\simeq100$ times larger than the bolometric flux, $F\simeq10^8\,{\rm L_\odot/kpc^2}$, implying an Eddington ratio of $\simeq0.01$. One can also make a crude estimate of the Eddington ratio in the optically-thin limit by using $F_{\rm Edd}\simeq2\pi G \Sigma_{\rm tot}c/\langle\kappa_{\rm RP}\rangle$.  The Eddington flux in the optically thin limit is $\simeq4\times10^9\,{\rm L_\odot/kpc^2}(\Sigma_{\rm gas}/30\,{\rm M_\odot/pc^2})(0.1/f_{\rm gas})(500\,{\rm cm^2/g}/\langle\kappa_{\rm RP}\rangle)$. We can compare this flux with the NUV, FUV, and optical fluxes. Averaged across the disk, these are of order a few times $10^7\,{\rm L_\odot/kpc^2}$, about 10 times smaller than the bolometric flux from the FIR emission. Again, we find that the system is substantially sub-Eddington when averaged over the face of the disk. Even assuming that $f_{\rm gas}\simeq1$ in the regions where most of the stars are forming the galaxy-averaged Eddington ratio is $0.02-0.1$. Yet, our results in this paper show that individual subregions may be super-Eddington, and perhaps strongly so, even though the galaxy-averaged measurements show that the system as a whole is not near the Eddington limit.

For a sample of high Eddington ratio regions we calculate the dynamics of the dusty material under simple assumptions, following the acceleration of the gas from small scales to larger scales,  including the effects of stellar aging as the dusty column is accelerated outward from the assumed central stellar cluster or the galaxy midplane (see Figure \ref{figure:velocity_plot}). Maximum velocities reach $10-20$\,km/s, even in the optimistic case that the shell does not sweep up gas as it expands.

As discussed in Sections \ref{section:spherical_case} and \ref{section:planar_case}, the inferred importance of radiation pressure can change significantly depending on our assumptions about the vertical scale along the line of sight of the dusty column and the geometry of the system. For a young stellar cluster, radiation pressure is a significant source of pressure on the gas in the immediate area, and for optically thin lines of sight this pressure is enough to accelerate the gas alone. However, as the material expands, the region rapidly loses its ability to support outward acceleration by radiation pressure alone due to three major effects: (1) the SED reddens as the population ages, (2) the bolometric luminosity decreases after $\gtrsim4$\,Myr, and (3) as the gas shell expands it encloses more old stellar mass that contributes significantly to the gravitational force. These effects preclude velocities above $\sim5-20$\,km/s in our sample, even for optimistic assumptions.

Tables \ref{table:GalaxyRegions} and \ref{table:GalaxyRegions10Myr} give  the parameters inferred for select super-Eddington regions in both galaxies assuming an age for the stellar population of 1 and 10\,Myr, respectively. In NGC 5194, region 37 presents an interesting case for discussion. The inferred Eddington ratio ranges from $11-0.1$, depending on the distance of the dusty gas to the star cluster ($H_{\rm gas}$). Assuming an age of 1\,Myr (Table \ref{table:GalaxyRegions}) the inferred new stellar mass $M_{\rm new,\,\star}$ is just 380\,M$_\odot$, which is insufficient to fully populate the IMF, and is thus inconsistent with our assumption that it is, as discussed in Section \ref{section:ObservationalQuantities}.However, because the ratio of the H$\alpha$ luminosity to the bolometric luminsoity is a strong function of age, at 10\,Myr (Table \ref{table:GalaxyRegions10Myr}) $M_{\rm new,\,\star}$ is $\simeq2.2\times10^4$\,M$_\odot$, plausibly fully-sampling the IMF, and range of Eddington changes to $\simeq4-0.4$, again depending on $H_{\rm gas}$. Regions 162 in NGC 5194 and region 27 in NGC 6296 have similar behavior. These examples help illustrate how the inferred Eddington ratio depends on age through the $L_{\rm H\alpha}/L_{\rm bol}$ ratio and its connection to $L_{\rm bol}/M_{\rm new\,\star}$.

For the purposes of our work here, we assume a single constant dust-to-gas mass ratio of $1/100$ for all galaxy regions, cloud radii, and ages. We further assume that the dust and gas are always dynamically coupled and uniformly mixed. Each of these may break down \citep{Hopkins2021}. Future works may include a variable $f_{\rm dg}$ ratio that tracks the gas-phase metallicity gradient of the host galaxy. Because $f_{\rm dg}$ is linearly related to the Eddington ratio, its role needs further exploration. 

Changing our assumed $f_{\rm dg}$  by a factor of 2 in either direction (e.g., to $1/50$ or $1/200$) roughly doubles (halves) the number of super-Eddington regions, with corresponding impact on the calculated velocities of super-Eddington regions. Additionally, we tested the impact of changing $f_{\rm dg}$ over time, to simulate the destruction of dust or sweeping up of additional dust into the dusty column as it is accelerated. Altering this ratio over time has little impact on the cloud dynamics, because most of the acceleration occurs early in the history of the region's evolution. Further work can be done to explore the importance of $f_{\rm dg}$ variations across galaxy sub-regions.

We note that adding the HI and CO gas measured in emission to the dusty gas measured by $A_{\rm H\alpha}$ in absorption may be double counting the total gas mass used to estimate the Eddington ratio. Because the cool/cold gas will have dust associated with it, and because that dust may contribute to the extinction of the region, we may under-estimate the Eddington ratio by counting each component separately. However, we find the Eddington ratio is not strongly dependent on HI and CO components, and is instead dominated by uncertainties in the age of the stellar population and the old stellar mass enclosed by the region, as determined here via the 3.6\,$\mu$m photometry. We thus conclude that potential double-counting in the gas mass is not a dominant uncertainty.  Additionally, under the simple assumptions employed here, the uncertainty in the dynamics of the super-Eddington regions is dominated by the assumed initial radius of the dusty column, the aging of the driving population, and the old stellar mass enclosed as the region expands. A more realistic treatment would include the dynamics of gas as it sweeps up more material on a region-by-region basis. This effect is not accounted for here, but would directly impact Equation \ref{equation:v_dvdr} as $M_{\rm gas}$ would become a function of distance from the driving region.

A limitation of the current work we highlight is the lack of stellar ages for each region.  Further, we do not use morphological information on each region, which might help constrain the height of the projected dusty gas column. Both of these quantities have a large impact on the outcomes in the sub-regions, changing the calculated bolometric luminosity, cluster mass, and enclosed mass. Figures \ref{figure:monte_carlo_result} and \ref{figure:WibkingComparison} show the decreasing momentum transfer efficiency for older stellar populations, due to the change in the SED of the cluster as it ages. Figures \ref{figure:Edd_hist}, \ref{figure:Edd_hist_planar}, and \ref{figure:Edd_ratio_Aha} and Tables \ref{table:GalaxyRegions} and \ref{table:GalaxyRegions10Myr} all show changes to the Eddington ratios from changing our base assumptions of age and column height.

The regions with the highest Eddington ratios tend to be optically thin, with $A_{\rm H\alpha} \sim 0.05$ (see Figure \ref{figure:Edd_ratio_Aha}). As expected for optically-thin super-Eddington regions, we find that a reasonable estimate of the maximum velocity reached by the expanding cloud in spherical geometry is 
\begin{equation}
v_{\rm max} \simeq v_{\rm esc}\sqrt{\frac{L_{\rm bol}}{L_{\rm Edd}}} \simeq \sqrt{\frac{2GM_{\rm tot}L_{\rm bol}}{H_{\rm gas}L_{\rm Edd}}}.
\end{equation}
Given that $L_{\rm bol}/L_{\rm Edd}$ can be as high as $10-50$ (see Fig.\ref{figure:KappaAndLOverM}), velocities are limited to $\sim10-20$\,km/s for the optically-thin sub-regions of the galaxies considered, given the minimum value of $H_{\rm gas}$ assumed (5\,pc), and for a typical value of the inferred total mass within those regions ($\sim10^3-10^4$\,M$_\odot$). For super-Eddington regions this means that the dusty gas will exceed the local escape velocity for the young cluster from the action of radiation pressure alone, prior to the effects of stellar aging. The outward expansion of the shell in this picture stops as it encloses more old stellar mass. One factor missed by this estimate is the impact of the radiation pressure flux and mass contributed from other sub-regions as the material expands and ``sees" more of the galaxy. Our estimates of these contributions suggest that they do not dominate the dynamics (Sections \ref{section:spherical_case} and \ref{section:modeling_cloud_velocities}).

Finally, we note that we have considered radiation pressure on dust alone as a feedback mechanism in the sub-regions considered. Radiation pressure may work together with stellar winds, proto-stellar jets, and other feedback processes in disrupting GMCs (e.g., \citealt{Murray2010a,Grudic2021}). In particular, recent work on the importance of the combined effects of massive star winds by \cite{Lancaster2021a} shows that the momentum input from stellar winds is similar to $L/c$, the momentum injection rate from photons in the single-scattering limit. If we were to add this piece to our Eddington ratios using the momentum injection rates for the stellar populations (e.g., using $[1+1-\exp(\langle\tau_{\rm RP}\rangle)](L/c)$), we find that more regions are super-Eddington and that the inferred velocities in the most super-Eddington regions are much larger. As an example, region 162 in NGC 5194 (see Tables \ref{table:GalaxyRegions} and \ref{table:GalaxyRegions10Myr}) can reach hundreds of km/s under the simplifying assumptions employed that the momentum input rate is constant at a value of $L/c$, independent of the projected column density and that there is no swept-up mass as the shell expands. 

In addition to the major numerical works underway to explore GMC disruption and feedback at the HII region scale (e.g., \citealt{Rathjen2021,Grudic2021}), future empirical and phenomenological assessments of feedback processes for large ensembles of sub-regions should include models for the variety of feedback mechanisms that have been proposed.

\appendix
\section{Dust Grain Distributions}
\label{appendix:DustGrains}

\begin{figure*}
    \centering
    \includegraphics[width=0.95\textwidth]{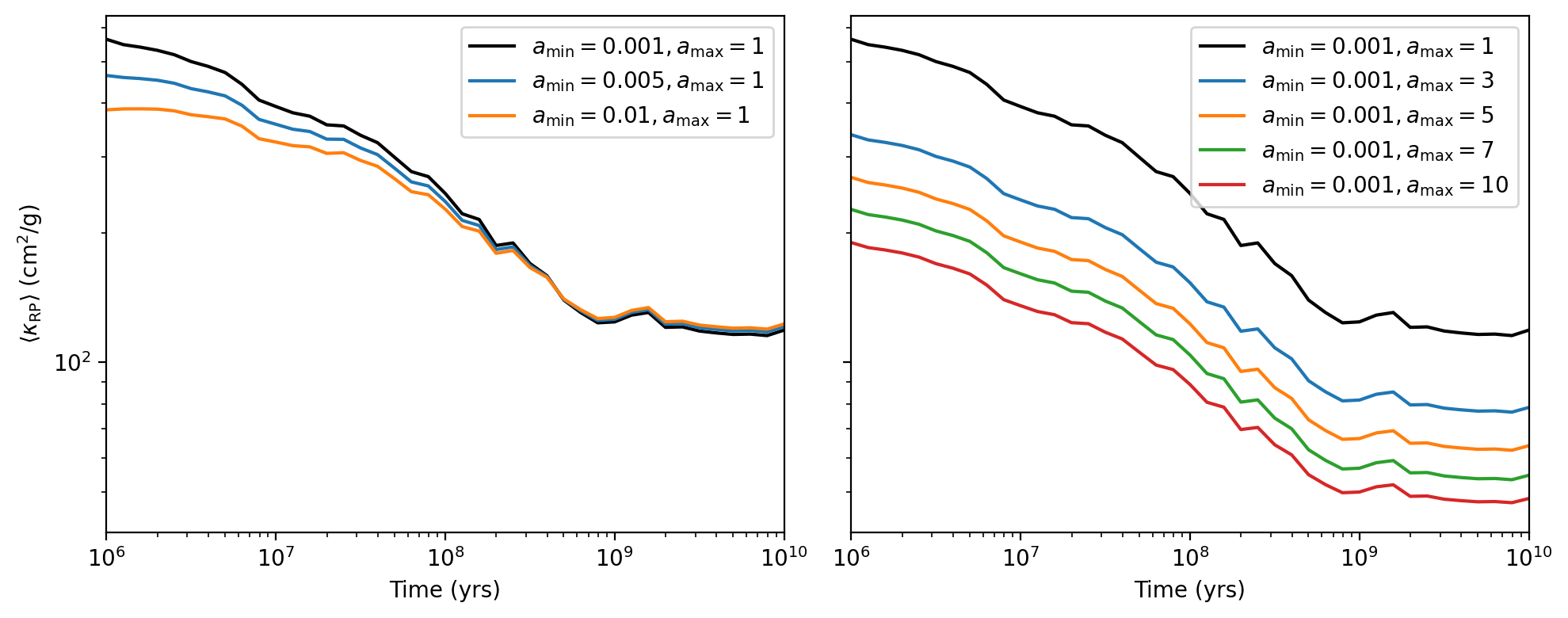}
    \caption{Values for $\langle\kappa_{\rm RP}\rangle$ with different choices of minimum grain size  $a_{\rm min}$ in $\mu$m (left) and maximum grain size $a_{\rm max}$ in $\mu$m (right) in the MRN distribution. Both panels assume our fiducial dust-to-gas mass ratio of $1/100$.}
    \label{figure:DustDistributionMaxAndMin}
\end{figure*}

In Section \ref{section:MonteCarloResults}, the effects of the dust grain distribution on the cloud opacity are discussed. The grain size distribution impacts the overall radiation pressure force, with different choices leading to an large difference in the resulting average opacity. Figure \ref{figure:grain_size_comparison} shows the values of $\langle\kappa_{\rm RP}\rangle$ over time for several different grain distributions. 

In order to explore these variations in more detail, in Figures \ref{figure:DustDistributionMaxAndMin} and \ref{figure:grain_size_comparison} we show the radiation pressure opacity from Equation \ref{equation:kappa_av_rp} under different choices for maximum and minimum grain sizes. The left and right panels in Figure \ref{figure:DustDistributionMaxAndMin}  show different choices for $a_{\rm min}$ and $a_{\rm max}$, respectively. As discussed in Section \ref{section:DustOpacitiesEddingtonRatios}, the opacity at all ages of the SED scales with $\approx 1/\sqrt{a_{\rm max}}$: for a fixed dust-to-gas mass ratio, a larger maximum grain size decrease the overall radiation pressure opacity. The minimum grain size affects the opacity most for young stellar populations when the SED still contains significant UV flux. At a fixed age of 1\,Myr with our fiducial BPASS SED, we find an approximate $\langle\kappa_{\rm RP}\rangle\propto a_{\rm min}^{-1/7}$.

This model neglects several effects that play an important role in the calculation of $\langle\kappa_{\rm RP}\rangle$ over time. First is that the assumption of a constant MRN distribution may be accurate for some dust grain ranges, but the shape of the distribution may not be MRN at the smallest grain sizes. Additionally, the distribution of each grain type in the mixture is assumed to be the same, however this may not be the case and the proportions of each grain type may change over the size distribution. Finally, the distribution may change over time, due to uneven destruction of different grain sizes. As discussed in Section \ref{section:discussion}, these changes over time are unlikely to have a large impact on the overall velocity achieved by a region under radiation pressure as the bulk of acceleration happens early in the dynamical expansion of a super-Eddington dusty cloud.

\section*{Acknowledgments}
\label{section:acknowledgements}
TAT acknowledges support from a Simons Foundation Fellowship in Theoretical Physics and an Einstein Fellowship from the Institute for Advanced Study, Princeton. This work was supported in part by National Science Foundation Grant \#1516967 and NASA ATP 80NSSC18K0526. We thank Sarah Kessler and Adam Leroy for sharing their datasets and for useful discussions. We thank Jiayi Sun for useful comments and discussions.

This work made use of v2.2.1 of the Binary Population and Spectral Synthesis (BPASS) models as described in \cite{Eldridge2017} and \cite{Stanway2018}. We also used several software libraries: Hoki, Numpy, Pandas, and Scipy \citep{Stevance2020,Harris2020,Reback2020,Mckinney2010,Virtanen2020}.

\section*{Data Availability}

The data underlying this article will be shared on reasonable request to the corresponding author.

\bibliographystyle{mnras}
\bibliography{bibliography}

\bsp	
\label{lastpage}
\end{document}